\documentclass[twocolumn]{aastex701}
\usepackage{tabularx}
\usepackage{amsmath}
\usepackage{amssymb}
\usepackage{longtable}
\usepackage{multirow}
\usepackage{bm}
\usepackage{mhchem}
\usepackage{rotating}
\usepackage{xcolor}

\begin{document}

\title{Toward Inferring the Surface Fluxes of Biosignature Gases on Rocky Exoplanets from Telescope Spectra}

\author[0000-0002-0413-3308]{Nicholas F. Wogan}
\affiliation{SETI Institute, Mountain View, CA 94043}
\affiliation{NASA Ames Research Center, Moffett Field, CA 94035}
\email{nicholas.f.wogan@nasa.gov}

\author[0000-0003-1240-6844]{Natasha E. Batalha}
\affiliation{NASA Ames Research Center, Moffett Field, CA 94035}
\email{natasha.e.batalha@nasa.gov}

\author{Joshua Krissansen-Totton}
\affiliation{Dept. Earth and Space Sciences / Astrobiology Program, University of Washington, Seattle, WA 98103}
\email{joshkt@uw.edu}

\author{Kevin Zahnle}
\affiliation{NASA Ames Research Center, Moffett Field, CA 94035}
\email{}

\author[0000-0002-1386-1710]{Victoria S. Meadows}
\affiliation{SETI Institute, Mountain View, California, 94043, USA}
\email{vmeadows@seti.org}

\author[0000-0003-3099-1506]{Amber V. Young}
\affiliation{Mary W. Jackson NASA Headquarters, Washington, DC, USA}
\email{}

\author[0000-0001-5290-1001]{Evan L. Sneed}
\affiliation{Department of Earth and Planetary Sciences, University of California, Riverside, CA 92521, USA}
\email{evan.sneed@email.ucr.edu}

\author[0000-0002-2949-2163]{Edward W. Schwieterman}
\affiliation{Department of Earth and Planetary Sciences, University of California, Riverside, CA 92521, USA}
\email{}

\begin{abstract}
  The James Webb Space Telescope and the future Habitable Worlds Observatory aim to discover exoplanet atmospheric spectra that detect life. Currently, most existing spectral ``retrieval'' algorithms focus on inferring the abundances of biogenic gases from these spectra. However, abundances are hard to interpret as signatures of life because they are modified by photochemistry, climate, and atmospheric escape. To address this problem, we develop a method for inferring the fluxes of gases at a planetary surface by inverting a coupled photochemical-climate model. As a proof-of-concept, we apply the approach to a synthetic 10-transit JWST NIRSpec Prism spectrum of TRAPPIST-1 e assuming it hosts a biosphere similar to the Archean Earth's. The retrieval confidently detects \ce{CO2} and \ce{CH4} and can constrain the flux of \ce{CH4} into the atmosphere to within approximately 1.5 orders of magnitude (68\% credible interval) provided that TRAPPIST-1's near-UV spectrum is accurately known. We demonstrate how inferred surface gas fluxes naturally fold into a probabilistic assessment of life, finding that $\sim80\%$ of the surface gas flux posterior is consistent with a \ce{CH4}-producing metabolism for our nominal test case. As with any inverse problem, these results are conditional on a number of assumptions in our forward model. Overall, we argue that increasing the robustness of life detection on exoplanets requires moving beyond atmospheric abundances toward inference of the surface fluxes that sustain them.
\end{abstract}

\section{Introduction}

Determining if exoplanets host life is a profound unsolved problem in science that has become a driving question for NASA. The James Webb Space Telescope (JWST) and large ground-based telescopes will search for biosignatures in a few favorable M dwarf systems with modest-resolution transit spectroscopy \citep{Meadows2023}, and high-resolution transit and reflected-light spectroscopy \citep{LopezMorales2019,Currie2025}. In the future, the space-based Habitable Worlds Observatory (HWO) will expand the search for signs of life on several dozen Earth-size planets orbiting Sun-like stars, using direct imaging and reflected-light observations \citep[e.g.,][]{KrissansenTotton2025,Arney2026,Parenteau2026}. The Large Interferometer For Exoplanets (LIFE), if realized, would provide a complementary pathway to biosignature detection via mid-infrared spectroscopy in a subsequent generation of space missions \citep{Quanz2022}.

Researchers can determine exoplanet properties and search for biosignatures from measured spectra with atmospheric retrieval methods \citep[e.g.,][]{MacDonald2017,Mukherjee2021,Molliere2019}. The goal of atmospheric retrieval is to construct a model of an exoplanet and its atmosphere that has a simulated spectrum that is consistent with measured data. The measured spectrum does not uniquely correspond to a single set of planetary parameters; therefore, a Bayesian approach is common where the researcher infers probability distributions of atmospheric properties (e.g. the O$_2$ abundance).

However, in the search for exoplanetary life, existing atmospheric retrieval models have two drawbacks. First, existing retrieval frameworks often permit models of exoplanet atmospheres that are not constrained by known photochemistry and climate physics. One example is O$_2$ and O$_3$, which are tightly coupled by photochemistry \citep{Seinfeld2016}. Many current retrieval frameworks do not account for the photochemical link between these gases, or any other gases (i.e., also called a ``free-chemistry'' retrieval). Accounting for such relationships during a retrieval would break degeneracies---a retrieved O$_3$ concentration implies likely O$_2$ abundances even if the spectrum does not contain detectable O$_2$ features \citep{Kozakis2022}. On the other hand, free-chemistry retrievals are valuable because they can reveal unexpected chemical or climate scenarios that are not captured by current models.

Perhaps the more important problem is that current retrieval methods determine the gas abundances that are compatible with spectra, but gas abundances alone are hard to interpret as a sign of life. Atmospheric abundances are determined by a complex interplay of photochemical sources and sinks, atmospheric escape, and surface fluxes, which includes the influence of biology if active \citep[e.g.,][]{KrissansenTotton2018a}. Inferring life or its absence should be much easier if we can estimate the fluxes of gas into the atmosphere at the planet's surface, because fluxes are closely linked to a biosphere's metabolic pathways. An example is the atmospheric methane biosignature \citep[e.g.,][]{Thompson2022,Guzman2013,Schindler2000}. It is unlikely that any single methane abundance value is indicative of life across all planetary systems, because its abundance is influenced by details of the incident UV spectrum, atmospheric temperature, and the redox state of the background atmosphere \citep{Segura2005,Wogan2020}. An estimated methane flux is easier to interpret because it can be directly compared to Earth's large biological flux and the relatively small plausible abiotic fluxes from geologic processes \citep{Wogan2020,KrissansenTotton2018a,KrissansenTotton2018b}, effectively normalizing out confounding planetary and astrophysical variables. 

To our knowledge, \citet{KrissansenTotton2018b} is the only study that explicitly attempted to infer a surface flux of a biosignature gas from a synthetic exoplanet spectrum. They performed a standard ``free-chemistry'' atmospheric retrieval on synthetic JWST NIRSpec Prism transmission spectra of an Archean Earth-like TRAPPIST-1 e finding that $\sim 10$ transits could detect the \ce{CH4}-\ce{CO2} disequilibrium biosignature \citep{KrissansenTotton2018a}. Next, they converted the inferred \ce{CH4} abundance to a minimum surface flux posterior by assuming \ce{CH4} is destroyed at least as rapidly as diffusion-limited hydrogen escape of the hydrogen atoms in the \ce{CH4}. However, \citet{Ranjan2025} used a photochemical model to show that this approximation is valid only when there are sufficient UV photons to destroy \ce{CH4} by either direct photolysis or indirectly by driving a catalytic cycle. This is not always the case for M-star hosted planets like TRAPPIST-1 e because M-stars can have a small UV output (see Section 4.1 of \citet{Ranjan2025} for details). Moreover, \citet{KrissansenTotton2018a} does not address how to infer the flux of biologically influenced gases beyond methane (e.g. O$_2$ or CO).

Here, we demonstrate a new atmospheric inference method that can more rigorously infer the surface flux of biogenic and other atmospheric gases from a planetary spectrum. We accomplish this by performing a grid-based Bayesian inversion of a coupled photochemical-climate model. The photochemical-climate model estimates the net production (or loss) of each atmospheric species from chemical reactions, condensation/evaporation, and escape; the result is the surface flux of a gas required to sustain a given atmospheric abundance. The model also naturally accounts for the photochemical relationships between gases (e.g., \ce{O2} and \ce{O3}). 

Following \citet{KrissansenTotton2018b}, we apply the new flux-inference method to synthetic JWST transmission spectra of TRAPPIST-1 e, assuming that it is an inhabited world similar to the Archean Earth, to assess the feasibility of inferring a biological methane flux on such a world. We choose TRAPPIST-1 e as a test case because it is one of the few habitable-zone rocky planets that are accessible with JWST and because the planet is being actively observed \citep{Allen2026}. However, our flux-inference methodology is applicable to any other potentially habitable world (e.g., LHS-1140 b) and to other spectroscopic datasets (emission, reflected-light, etc.). This study is intended as a proof-of-concept demonstration of flux retrieval and does not attempt to marginalize over all uncertainties in the forward model. Finally, using TRAPPIST-1 e as an example, we demonstrate how an inferred surface flux of a biosignature gas readily permits a probabilistic inference of exoplanetary life.

\section{Methods} \label{sec:method}

\subsection{Forward-model overview}

\begin{figure*}
  \centering
  \includegraphics[width=\textwidth]{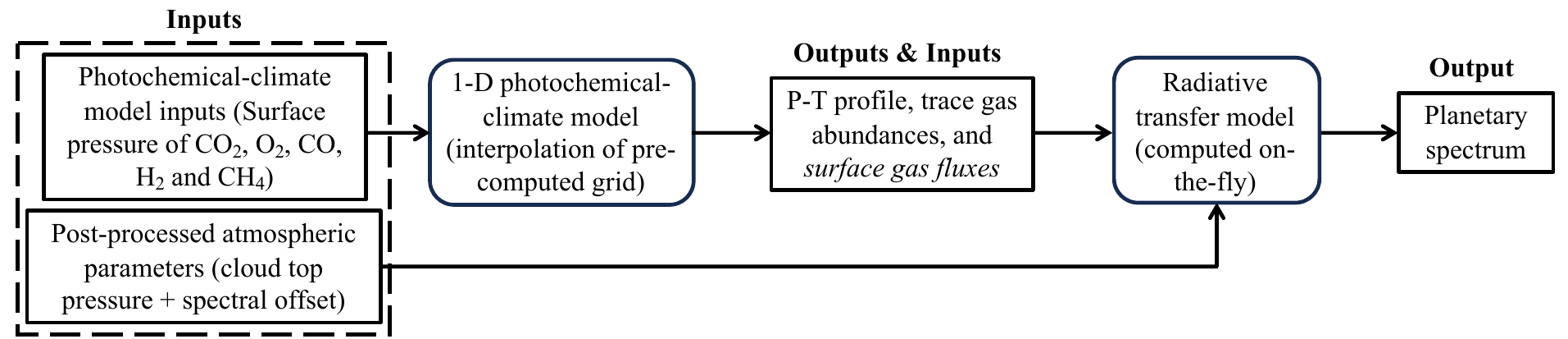}
  \caption{A schematic of the forward model that we use to retrieve surface gas fluxes and other atmospheric parameters from telescope spectra. The inputs include the surface partial pressure of bulk gases (\ce{CO2}, \ce{O2}, \ce{CH4}, etc.), and cloud parameters. With the surface gas partial pressures the model linearly interpolates a pre-computed grid of 1-D photochemical-climate simulations to predict the pressure-temperature (P-T) profile, trace gas abundances (\ce{C2H6}, \ce{O3}, etc.) and surface gas fluxes. This predicted atmosphere, along with the input cloud parameters, are fed into a radiative transfer code to generate a planetary spectrum.}
  \label{fig:fm}
\end{figure*}

Figure \ref{fig:fm} shows our forward model that we invert to infer the surface flux of biosignature gases on habitable-zone rocky exoplanets. The inputs are split into two categories. The first category includes parameters that are inputs to a one-dimensional photochemical-climate atmospheric model (see Section \ref{sec:method_pc}). For our TRAPPIST-1 e test case, these inputs are the surface partial pressures of \ce{CO2}, \ce{O2}, CO, \ce{H2} and CH$_4$ (all in $\log_{10}$ bars), although additional variables could also be considered (see Section \ref{sec:discussion_caveat} discussion). Given these fixed partial pressures at the lower boundary, the photochemical-climate model predicts the pressure-temperature profile (i.e., P-T profile), the gas abundances above the surface (including trace gases like \ce{C2H2}, \ce{O3}, etc.), and the surface gas fluxes of the fixed gases. The photochemical-climate model is relatively slow ($\sim5$ core-minute per simulation), so we pre-compute a grid of atmospheres and linearly interpolate the resulting compositions and climates during Bayesian inversion.

The atmosphere predicted by the photochemical-climate model is an input to a radiative transfer code (Section \ref{sec:method_rt}) that computes a planetary spectrum. The radiative transfer code also accounts for the second category of inputs to our forward model (Figure \ref{fig:fm}): ``post-processed'' atmospheric parameters. In our TRAPPIST-1 e test case, we have two such variables: a gray cloud top pressure (in $\log_{10}$ bars) and a spectral offset to marginalize over planet radii. We set the cloud optical depth to 10 to rapidly attenuate radiation entering the cloud layer. The simple cloud parameterization is meant to represent any possible atmospheric aerosol layer (water cloud or haze).

In the following subsections, we describe each component of the Figure \ref{fig:fm} forward model in the context of our TRAPPIST-1 e test scenario. Section \ref{sec:method_pc} addresses the photochemical-climate model and corresponding grid of calculations, while Section \ref{sec:method_rt} describes the radiative transfer code. Section \ref{sec:method_case} then details the TRAPPIST-1 e case study to which we apply our flux-inference method, and Section \ref{sec:methods_interp} defines the flux-based biosignature framework used later for probabilistic interpretation.

\subsection{The photochemical-climate model} \label{sec:method_pc}

\subsubsection{Model description}

Our flux-inference method incorporates the one-dimensional (1-D) photochemical and climate models in the \texttt{Photochem} software package \citep[version 0.6.8;][]{Photochem2025}. \citet{Wogan2025b} describes the model in full, and validates the code against the observed composition and climates of Venus, Earth, Mars, Jupiter and Titan. Here we provide a brief overview of the code.

The climate model solves for radiative and convective energy balance in the atmosphere. To estimate the radiation field, the code uses standard two-stream methods \citep{Toon1989} with opacities representing photolysis, Rayleigh scattering, collision-induced absorption (CIA) and approximates line absorption with k-distributions \citep{Amundsen2017}. In this work, we include opacities for \ce{H2}, \ce{H2O}, \ce{O2}, \ce{CO}, \ce{CO2}, \ce{CH4} and \ce{N2}. Most opacities are derived from HITEMP or HITRAN \citep{Gordon2017,Hargreaves2020}, although \citet{Wogan2025b} provides a complete description of each opacity and their origin in the literature. Convection occurs in the model if the temperature gradient exceeds a moist or dry pseudo-adiabat \citep{Graham2021}. The model iterates to radiative-convective equilibrium directly with a variation of Newton's method \citep{Wogan2025b}.

The photochemical model solves for the steady-state of a system of partial differential equations describing how atmospheric gases are influenced by chemical reactions, condensation/evaporation, escape, and vertical transport \citep[Equation 1 in][]{Wogan2025b}. Here, we use the chemical network and thermodynamics of \citet{Wogan2025b} limited to species composed of H, N, O and C. In total, the model tracks 73 gas-phase and 14 condensed-phase species modulated by 483 reversible chemical reactions and 73 non-reversible photolysis reactions. The network includes a routine for estimating the production of hydrocarbon haze, which has been benchmarked against Titan's observed haze production rate \citep{Wogan2025b}. The haze only serves to remove hydrocarbons from the atmosphere by sedimentation, and we do not include the predicted haze profile when generating planetary spectra (and instead adopt a parameterized gray cloud).

We couple the climate and photochemical models to generate P-T and composition profiles: first, given the input partial pressures of bulk gases (e.g., \ce{N2}, \ce{CO2}, \ce{CO}, \ce{O2}, \ce{CH4}, \ce{H2O} and \ce{H2}), we compute a climate at radiative-convective equilibrium. Next, we fix this resulting P-T profile and corresponding bulk surface partial pressures in the photochemical model and compute a photochemical steady-state atmospheric composition. We do not perform further iterations back and forth between the climate and photochemical models to simulate a perfectly self-consistent composition and climate. We choose a loose coupling to maintain computational efficiency and stability over the large grid of calculations required to infer surface gas fluxes. Appendix \ref{sec:appendix_couple} shows that this approach is a reasonable approximation, especially for an Archean Earth-like atmosphere.

\subsubsection{Grid of TRAPPIST-1 e calculations} \label{sec:method_grid}

For our TRAPPIST-1 e case study, we generate a grid of photochemical-climate simulations that we linearly interpolate over the course of a retrieval. Specifically, the grid covers all the Table \ref{tab:params} surface partial pressures combinations for \ce{CO2}, \ce{O2}, \ce{CO}, \ce{H2} and \ce{CH4}, totaling 34,992 simulations. These model runs share several assumptions. In all simulations, we assume 1 bar of \ce{N2}, a surface reservoir of \ce{H2O} (i.e., an ocean), and impose a 50\% relative humidity of condensation for all gases (although \ce{H2O} is the main condensate). Furthermore, the surface albedo is set to 0.1 and the eddy diffusion coefficient is fixed to a vertically-constant $10^5$ cm$^2$ s$^{-1}$. The photochemical model adopts the deposition velocities in Table 6 of \citet{Ranjan2023}. H and \ce{H2} escape the atmosphere at the diffusion-limited rate. We use the planetary and stellar parameters (mass, radius, bolometric insolation) derived in \citet{Agol2021}. We nominally adopt the HAZMAT spectrum for TRAPPIST-1 from \citet{Peacock2019} (their ``model group 1a''), although we perform a sensitivity test with the MUSCLES TRAPPIST-1 semi-empirical spectrum from \citep{Wilson2021}.

\begin{table}
  \begin{center}
  \caption{Parameters in the photochemical \& climate grid} \label{tab:params}
  \begin{tabular}{p{0.25\linewidth} p{0.64\linewidth}}
  \hline \hline
  Parameter & Values ($\log_{10}$ bars) \\
  \hline
  $\log_{10}(P_\mathrm{CO_2})$$^\dagger$ & -9, -7, -5, -4, -3, -2, -1, 0, 1 \\
  $\log_{10}(P_\mathrm{O_2})$ & -19, -15, -11, -7, -5, -3, -1, 1 \\
  $\log_{10}(P_\mathrm{CO})$ & -11, -7, -5, -3, -1, 1 \\
  $\log_{10}(P_\mathrm{H_2})$ & -10, -8, -6, -4, -2, 0 \\
  $\log_{10}(P_\mathrm{CH_4})$ & -13, -11, -9, -7, -6, -5, -4, -3, -2, -1, 0, 1 \\
  \hline
  \multicolumn{2}{p{0.97\linewidth}}{$^\dagger$A fixed surface pressure in $\log_{10}$ bars.}\\
  \end{tabular}
  \end{center}
\end{table}

We emphasize that the grid, and consequently our inversion for surface fluxes, necessarily assumes several planetary properties that may not be constrained by JWST observations alone, including the vertical mixing coefficient, the presence of a surface ocean, and 1 bar of \ce{N_2}. Ideally, the grid would cover a wider range of uncertain parameters although the ``curse of dimensionality'' limited us to the Table \ref{tab:params} grid without expending vast amounts of computational resources. Section \ref{sec:discussion_caveat} discusses the consequences of failing to capture all the uncertainty in the photochemical-climate model.



\begin{figure*}[t!]
  \centering
  \includegraphics[width=0.9\textwidth]{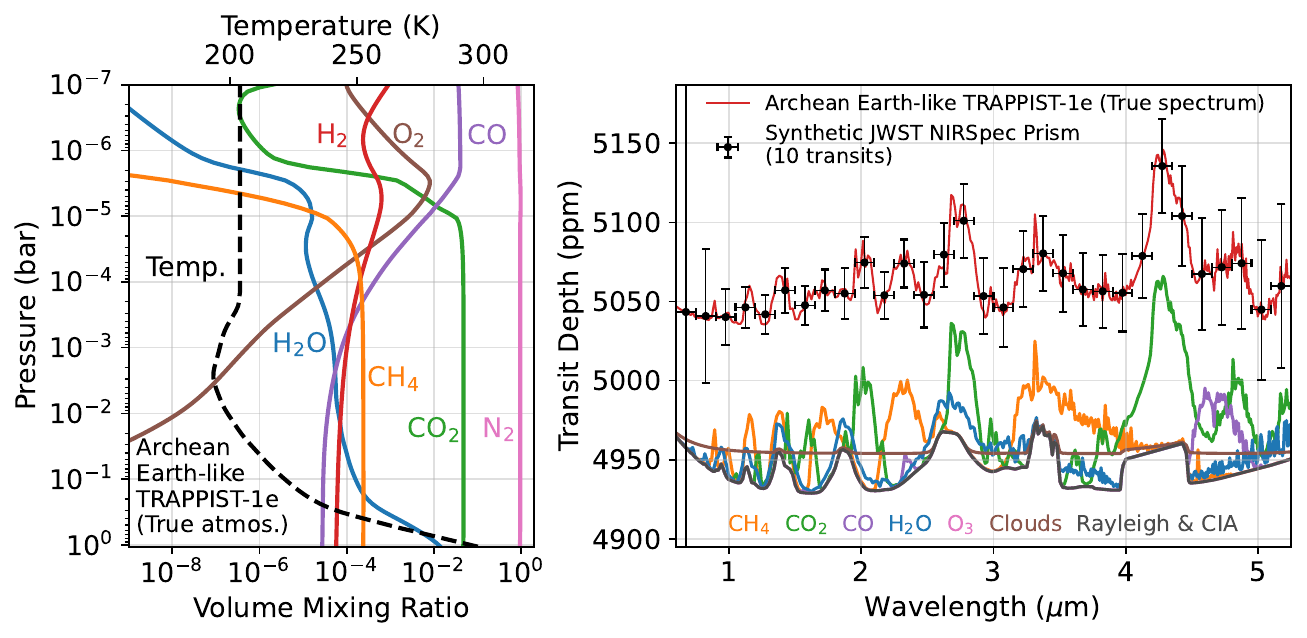}
  \caption{A simulated Archean Earth-like TRAPPIST-1 e, to which we apply our novel retrieval algorithm. Left: The modeled temperature profile (black dashed line) and mixing ratio profiles (solid color lines) of main atmospheric species for an Archean Earth-like TRAPPIST-1 e. The simulation has a biological CH$_4$ flux equal to the modern Earth's resulting in a 234 ppmv surface abundance. Right: The transmission spectrum of the Archean Earth-like atmosphere (red line). The data points are a synthetic NIRSpec Prism spectrum for 10 transits computed with \texttt{PandExo} assuming no stellar contamination. The colored lines below the data indicate the spectral contribution of various molecules, clouds, Rayleigh scattering, and CIA, all shifted downward for visual clarity.}
  \label{fig:true_atmos}
\end{figure*}

\subsection{Planetary spectra} \label{sec:method_rt}

We use the \texttt{PICASO} code \citep{Batalha2019} to compute transmission spectra of TRAPPIST-1 e. We use a set of $R = \mathrm{15,000}$ resampled opacities archived on Zenodo \citep{Opacities2025}, derived from a similar set of HITEMP and HITRAN opacities used by \texttt{Photochem} for climate calculations (Section \ref{sec:method_pc}). The opacity database accounts for line absorption from the following relevant species: \ce{C2H2}, \ce{C2H6}, \ce{CH4}, \ce{CO}, \ce{CO2}, \ce{H2O}, \ce{NH3}, \ce{O2}, \ce{O3}. Also, relevant CIA partners include \ce{CH4}-\ce{CH4}, \ce{CO2}-\ce{CH4}, \ce{CO2}-\ce{CO2}, \ce{CO2}-\ce{H2}, \ce{H2}-\ce{CH4}, \ce{H2}-\ce{H2}, \ce{H2O}-\ce{H2O}, \ce{H2O}-\ce{N2}, \ce{N2}-\ce{CH4}, \ce{N2}-\ce{H2}, \ce{N2}-\ce{N2}, \ce{N2}-\ce{O2}, and \ce{O2}-\ce{O2}. Here, the water continuum \citep[MT\_CKD v3.5;][]{Mlawer2012} is reformulated as a \ce{H2O}-\ce{H2O} (self) and \ce{H2O}-\ce{N2} (foreign) contribution. \texttt{PICASO} also folds in Rayleigh scattering for all specified molecules. We include a single gray and opaque cloud deck ($\tau = 10$) that extends from the surface to a specified pressure level. Note that clouds only affect the transmission spectrum -- the atmosphere is assumed to be largely cloud free for radiative transfer calculations in the 1-D photochemistry and climate.

\subsection{A TRAPPIST-1 e flux-inference case study} \label{sec:method_case}

\subsubsection{The ``true'' Archean Earth-like TRAPPIST-1 e atmosphere}

We test the flux-inference method on a synthetic JWST NIRSpec Prism transmission spectrum of TRAPPIST-1 e assuming the planet has an atmosphere and biosphere like the Archean Earth's. Specifically, we assume TRAPPIST-1 e has a 1 bar \ce{N2}-dominated atmosphere with $\sim5\%$ \ce{CO2}, to increase the likelihood of a habitable surface temperature \citep{Meadows2023}, influenced by \ce{CH4}-producing and CO-consuming microbes. We represent methanogenic life ($\ce{CO2 + 4 H2 -> CH4 + 2 H2O}$) with a surface \ce{CH4} flux into the atmosphere identical to Modern Earth's \citep[$1.122 \times 10^{11}$ molecules cm$^{-2}$ s$^{-1}$;][]{Catling2017}, and an \ce{H2} deposition velocity of $2.4 \times 10^{-4}$ cm s$^{-1}$ following \citet{Kharecha2005}. Also, based on \citet{Kharecha2005}, we adopt a CO deposition velocity of $1.2 \times 10^{-4}$ cm s$^{-1}$ to model the influence of a CO-consuming metabolism ($\ce{4 CO + 2 H2O -> 2 CO2 + CH3COOH}$). Implicitly, this approach also implies acetotrophic methanogens that convert \ce{CH3COOH} to \ce{CH4} and \ce{CO2} \citep{Kharecha2005}. We do not consider any volcanic surface fluxes of \ce{H2} or CO. This is a valid assumption for Earth-like outgassing, which has small fluxes of \ce{H2} and CO ($\leq3.7 \times 10^9$ molecules cm$^{-2}$ s$^{-1}$; \citet{Catling2017}), compared to the large rate at which \ce{H2} and CO are sequestered into the planet by biology in our nominal model ($\geq 8\times10^{10}$ molecules cm$^{-2}$ s$^{-1}$). Otherwise, our model adopts the other parameters described in the grid of TRAPPIST-1 e simulations (Section \ref{sec:method_grid}), like the deposition velocities in Table 6 of \citet{Ranjan2023}, a 0.1 surface albedo, and a vertically constant eddy diffusion of $10^5$ cm$^2$ s$^{-1}$.

Figure \ref{fig:true_atmos}, left, shows the atmospheric temperature and composition predicted by the forward model for these inputs. The flux of \ce{CH4} from methanogens sustains 234 ppmv \ce{CH4} at the surface against photochemical destruction, while biological deposition of CO and \ce{H2} result in 27 and 57 ppmv, respectively. The \ce{CO2} and \ce{CH4} greenhouse supports a habitable 298 K surface. As a post-processing step to our photochemical-climate calculation, we impose a gray and opaque cloud that extends from the surface to 0.2 bar as this is the approximate region of \ce{H2O} condensation. Figure \ref{fig:true_atmos}, right, illustrates the transmission spectrum of the planet, along with each molecule's contribution. \ce{CH4} and \ce{CO2} features dominate and there is minor CO absorption near 4.8 $\mu$m. Overall, the atmospheric composition and spectrum is similar to the ``Archean warm (cloudy)'' case modeled in \citet{Meadows2023}.

\subsubsection{Synthetic JWST NIRSpec Prism observations}

We use \texttt{PandExo} \citep{Batalha2017} to compute a synthetic 10 transit JWST Prism spectrum for the modeled Archean Earth-like TRAPPIST-1 e. Figure \ref{fig:true_atmos}, right, shows the data with 0.15 $\mu$m bins for visual clarity, although our Bayesian inversions (described below) use native NIRSpec Prism resolution. We center the synthetic observations on our Archean Earth-like scenario, ignoring random Gaussian noise to avoid an anomalous noise realization. In reality, achieving the precision of this synthetic data would likely require more than 10 transits as \texttt{PandExo} predicts errors $\sim 20\%$ smaller than real-world spectra \citep[e.g.,][]{Scarsdale2024}. Also, extracting a planetary spectrum will require techniques for eliminating stellar contamination \citep[e.g.,][]{Rathcke2025} given the strong contamination in the first four published TRAPPIST-1 e transits \citep{Espinoza2025}. Nevertheless, these real-world challenges do not affect the validity of this proof-of-concept demonstration of flux inference from spectra.

\subsubsection{Bayesian inversion for surface fluxes and other parameters}

Given the synthetic JWST TRAPPIST-1 e spectrum in Figure \ref{fig:true_atmos}, right, we do a Bayesian inversion of our forward model (Figure \ref{fig:fm}) using the \texttt{PyMultiNest} algorithm \citep{Buchner2014} adopting 1000 live points and an evidence tolerance of 0.5 to ensure convergence. We infer seven total parameters: five of the parameters are the surface partial pressures of \ce{CO2}, \ce{O2}, \ce{CO}, \ce{H2}, and \ce{CH4} in $\log_{10}$ space. For each gas, we adopt uniform priors that span the range of values in our pre-computed grid (Table \ref{tab:params}; e.g., \ce{CO2} has a uniform prior between -9 and 1 $\log_{10}$ bars). The sixth parameter is a gray and opaque cloud top pressure with a uniform prior between -5 and 0 $\log_{10}$ bars. Finally, we fit for a vertical shift to the transmission spectrum (to accommodate for uncertain planet radius and reference pressure) with a uniform prior between -1000 and 1000 ppm. Each set of partial pressures (\ce{CO2}, \ce{O2}, \ce{CO}, \ce{H2}, \ce{CH4}) corresponds to a set of surface gas fluxes necessary to sustain each gas abundance. Therefore, as a post-processing step, we derive posteriors for surface gas fluxes and other atmospheric properties (e.g., volume mixing ratios) by feeding samples from the partial pressure posteriors through our forward model.

While our Bayesian inversion adopts explicit log-uniform priors on surface partial pressures, the mapping between partial pressure and sustaining surface flux in our forward model induces non-uniform implicit priors on the corresponding fluxes. Appendix Figure \ref{fig:corner_flux} shows the implicit priors on each surface gas flux. When defining the bounds of our pre-computed grid (Table \ref{tab:params}), we ensured that the induced implicit priors on surface fluxes did not artificially constrain the posterior distributions. For example, the grid extends to $P_\mathrm{O_2} = 10^{-19}$ bar so that the resulting prior on the \ce{O2} surface flux is approximately symmetric between net sources and net sinks (i.e., fluxes into and out of the atmosphere). Methane is hard to synthesize photochemically \citep[e.g.,][]{Wogan2024}, so the implicit flux prior has some preference for positive values even though we consider \ce{CH4} surface pressures down to $10^{-13}$ bar.

\subsection{Flux-based biosignature definitions} \label{sec:methods_interp}

To interpret inferred surface gas fluxes in our TRAPPIST-1 e case study as evidence for an inhabited or lifeless planet, we must define how those fluxes map to biosignatures. Therefore, in this section we define biosignatures and false positives directly in terms of surface fluxes using a set-based framework. Later, in Section \ref{sec:results_interp}, we apply these definitions to our TRAPPIST-1 e flux posteriors to compute the probability that the inferred fluxes are consistent with life.

Let $\mathbf{F} = (F_{\mathrm{CH}_4}, F_{\mathrm{CO}}, F_{\mathrm{H}_2}, F_{\mathrm{O}_2}, F_{\mathrm{CO}_2})$ denote the vector of surface gas fluxes, and let $\mathcal{F}$ denote the space of all such flux vectors allowed by our retrieval priors for these gases. Our goal is to define a subset of $\mathcal{F}$ that corresponds to metabolism influencing a planet. Here, we develop a simple and interpretable definition of this ``biosignature space'', although the same set-based framework can accommodate more sophisticated definitions.

First, we describe the subset of $\mathcal{F}$ corresponding to an Archean Earth-like \ce{CH4} biosignature. A surface \ce{CH4} flux is a compelling biosignature when the flux is too large to explain with non-living processes, and thus can be best explained by efficient methanogenic microbes. So, establishing a \ce{CH4} biosignature requires estimates for how rapidly the gas can be produced by abiotic processes. \citet{KrissansenTotton2018a} estimated an upper-bound for abiotic \ce{CH4} production from ocean-bottom serpentinization, where iron minerals in the seafloor react with water to produce \ce{H2} which subsequently reduces \ce{CO2}. They compute a probability distribution for this upper bound; however, for simplicity, we adopt a single upper limit for abiotic serpentinization of $3.7\times10^{10}$ molecules cm$^{-2}$ s$^{-1}$ (or 10 Tmol/yr on Earth), corresponding to the $\sim 99.7^{\mathrm{th}}$ percentile of their distribution. \citet{Thompson2022} found that other upper-bound literature estimates for methane production from water-rock or metamorphic reactions on Earth over time are all $< 3.7\times10^{10}$ molecules cm$^{-2}$ s$^{-1}$, meaning our adopted upper limit is very conservative. We define the subset of flux space corresponding to methane fluxes too large to be explained by abiotic serpentinization:
\begin{equation} \label{eq:def_h_ch4}
\mathcal{H}_{\mathrm{CH}_4}
=
\left\{
\mathbf{F} \mid F_{\mathrm{CH}_4} > 3.7\times10^{10}
\right\}.
\end{equation}
In this set notation, the expression on the right-hand side should be read as ``the set of all flux vectors $\mathbf{F}$ for which the surface \ce{CH4} flux exceeds $3.7\times10^{10}$ molecules cm$^{-2}$ s$^{-1}$.'' \cite{Wogan2020} argued that volcanism may still be able to produce large \ce{CH4} fluxes within the space $\mathcal{H}_{\mathrm{CH}_4}$, however, such reducing outgassing would also likely generate more \ce{CO} than \ce{CH4} and significant \ce{H2}. Therefore, we define the ``reducing volcanic false-positive'' subset within $\mathcal{H}_{\mathrm{CH}_4}$ as,
\begin{equation} \label{eq:def_v}
\mathcal{V}
=
\left\{
\mathbf{F}\in\mathcal{H}_{\mathrm{CH}_4}
\mid
F_{\mathrm{CO,max}}^\uparrow > F_{\mathrm{CH}_4}
\;\wedge\;
F_{\mathrm{H_2,max}}^\uparrow > 0
\right\}.
\end{equation}
Here, the symbol $\wedge$ denotes a logical ``and'', so that $\mathcal{V}$ consists of flux vectors in $\mathcal{H}_{\mathrm{CH}_4}$ for which $F_{\mathrm{CO,max}}^\uparrow > F_{\mathrm{CH}_4}$ and $F_{\mathrm{H_2,max}}^\uparrow > 0$. The quantities $F_{\mathrm{CO,max}}^\uparrow(\mathbf{F})$ and $F_{\mathrm{H_2,max}}^\uparrow(\mathbf{F})$ are maximum upward fluxes of \ce{CO} and \ce{H2}, respectively, implied by the net flux vector $\mathbf{F}$ under a deposition-limited uptake model, as derived in Appendix \ref{sec:appendix_deriv1} following \citet{Kharecha2005}. A planet could have a large flux of both \ce{CO} and \ce{H2} into the atmosphere from volcanism, as well as a large surface deposition into the planet, with net fluxes close to zero. Therefore, Equation \eqref{eq:def_v} uses estimated maximum upward fluxes (e.g., $F_{\mathrm{CO,max}}^\uparrow$) instead of the net surface flux (e.g., $F_{\mathrm{CO}}$).

In sum, we define an Archean Earth-like methane biosignature as,
\begin{equation} \label{eq:def_b_ch4}
\mathcal{B}_{\mathrm{CH}_4}
=
\mathcal{H}_{\mathrm{CH}_4} \setminus \mathcal{V}
\end{equation}
Here, $\setminus$ denotes set subtraction, so that $\mathcal{B}_{\mathrm{CH}_4}$ is the space where $F_{\mathrm{CH}_4}$ is too large to be explained by serpentinization ($\mathcal{H}_{\mathrm{CH}_4}$), but excludes the reducing volcanic false-positive ($\mathcal{V}$). Furthermore, we also define a Modern Earth-like oxygen biosignature as the subset of $\mathcal{H}_{\mathrm{CH}_4}$ with a net positive \ce{O2} flux into the atmosphere:
\begin{equation} \label{eq:def_b_o2}
\mathcal{B}_{\mathrm{O}_2}
=
\left\{
\mathbf{F}\in\mathcal{H}_{\mathrm{CH}_4}
\mid
F_{\mathrm{O}_2} > 0
\right\}.
\end{equation}
The overall region of flux space consistent with either an Archean Earth-like or Modern Earth-like biosignature is the union.
\begin{equation} \label{eq:def_l}
\mathcal{L}
=
\mathcal{B}_{\mathrm{CH}_4} \cup \mathcal{B}_{\mathrm{O}_2}.
\end{equation}
We also assume that all fluxes outside $\mathcal{L}$ are an abiotic space that can be explained by non-living processes:
\begin{equation} \label{eq:def_a}
\mathcal{A} = \mathcal{F} \setminus \mathcal{L}
\end{equation}

Our definitions of the biosignature space (i.e., $\mathcal{L}$) and abiotic space (i.e., $\mathcal{A}$) in terms of surface gas fluxes are intentionally simple and discrete in order to illustrate the framework. In reality, the gas fluxes corresponding to inhabited and uninhabited worlds likely overlap, and the distinction between biological and abiotic regimes is inherently probabilistic rather than binary. In such a generalized formulation, one would replace the sharp set $\mathcal{L}$ with a probability density or weighting function $p(\mathrm{life}\mid\mathbf{F})$, which assigns a higher probability of life to some regions of flux space and lower probability to others. The discrete sets defined here can therefore be interpreted as a limiting case of this more general framework, in which $p(\mathrm{life}\mid\mathbf{F})$ is approximated as either zero or one. 

\section{Results} \label{sec:results}

\subsection{Inferred atmospheric abundances}

\begin{figure*}[t!]
  \centering
  \includegraphics[width=0.85\textwidth]{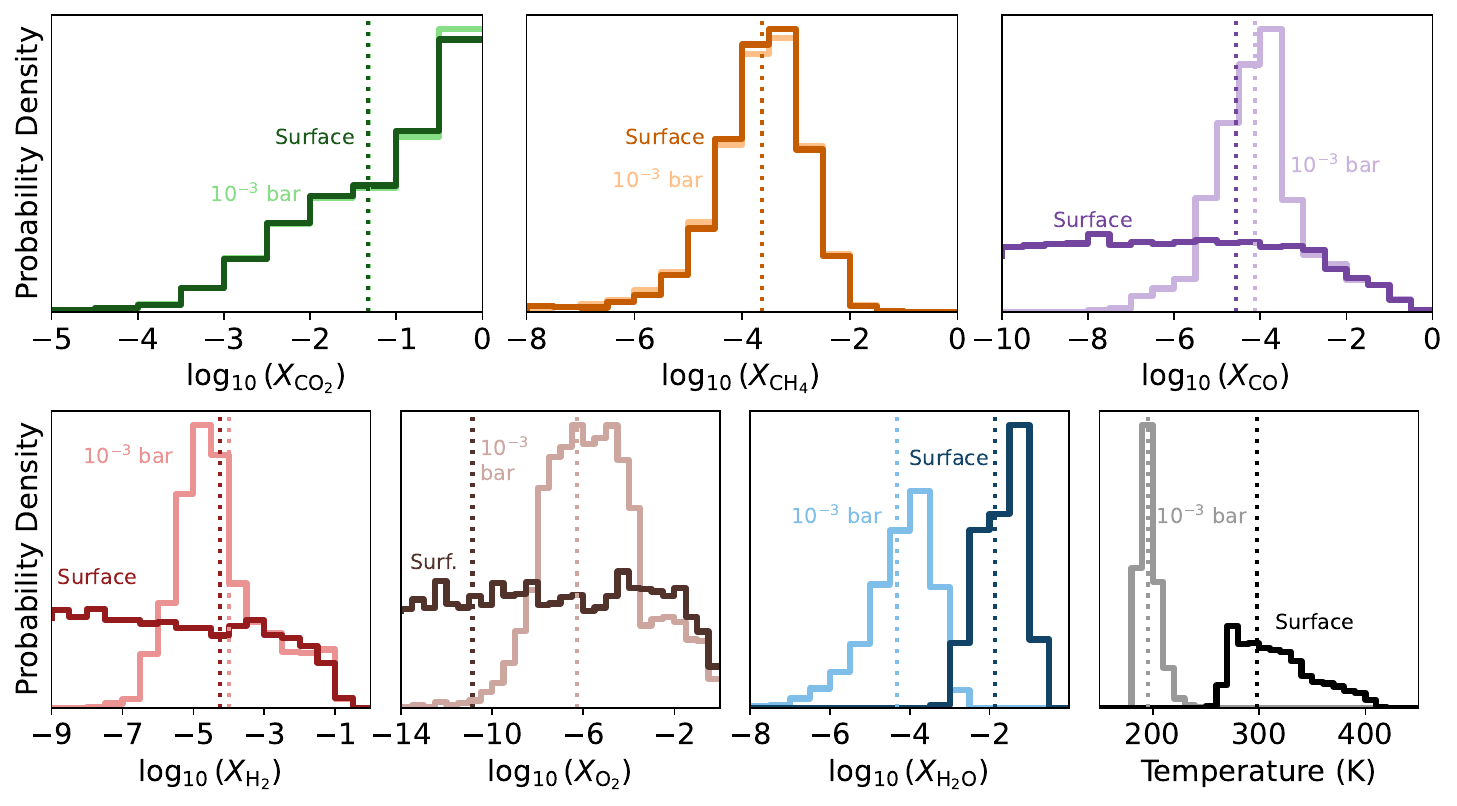}
  \caption{Inferred atmospheric volume mixing ratios and temperatures for an Archean Earth-like TRAPPIST-1 e observed with 10 JWST Prism transits (Figure \ref{fig:true_atmos}). Each panel has two posterior distributions: one for the surface atmospheric layer and another for the $10^{-3}$ bar layer of the atmosphere. The vertical dotted lines are the true values. The retrieval places tight constraints on \ce{CO2} and \ce{CH4} because the synthetic spectrum is sensitive to both molecules (Figure \ref{fig:true_atmos}). The retrieval has peaked posteriors for CO, O$_2$ and H$_2$ at $10^{-3}$ bar because they are photochemical byproducts of the detected CO$_2$ and CH$_4$.}
  \label{fig:abundances}
\end{figure*}


Figure \ref{fig:abundances} shows the surface and $10^{-3}$ bar mixing ratios and temperatures inferred by our inverse method from the synthetic JWST TRAPPIST-1 e spectrum in Figure \ref{fig:true_atmos}. While our inversion constrains surface partial pressures, we derive mixing ratio profiles by propagating the posterior samples through the forward model (Figure \ref{fig:fm}). The \ce{CO2} and \ce{CH4} posteriors show clear detections because the spectrum is sensitive to both molecules. In contrast, the surface abundances of \ce{CO}, \ce{H2} and \ce{O2} are poorly constrained, only ruling out concentrations larger than -1.9, -1.6 and -1.4 $\log_{10}$ mixing ratio (95\% credible), respectively. These surface abundance constraints are broadly consistent with previous free retrieval studies of an Archean TRAPPIST-1 e \citep{KrissansenTotton2018b,MikalEvans2022}. 

Unlike the surface abundances, the $10^{-3}$ bar posteriors in Figure \ref{fig:abundances} for CO, \ce{H2} and \ce{O2} are peaked. The constraints do not come directly from spectral features, as these molecules do not have strong absorption in the JWST Prism band pass (Figure \ref{fig:true_atmos}). Instead, the inversion, which embeds a photochemical model, is sensitive to stratospheric CO, \ce{H2}, and \ce{O2} because they are photochemical byproducts of the strongly detected \ce{CO2} and \ce{CH4}.

The transmission spectrum does not constrain the upper-bound surface partial pressure of \ce{CO2} (Appendix Figure \ref{fig:corner}), permitting up to 4.8 bar \ce{CO2} (95\% credible). As a result, the $1\sigma$ constraint on surface temperature is 279 to 357 K (Figure \ref{fig:abundances}), biased to higher temperatures because of the possibility of a strong \ce{CO2} greenhouse. The confident $\lesssim 1\%$ upper bound on \ce{CH4} (Figure \ref{fig:abundances}) contributes to the tight constraints on stratospheric temperature (188 to 205 K, 68\% CI). Without large \ce{CH4} concentrations, the atmosphere has few molecules capable of heating the stratosphere by starlight absorption, so the upper atmosphere tends towards the planet's 192 K skin temperature \citep[for a bond albedo of 0.3;][]{Catling2017}. The inferred temperature profile, along with our assumption of a large surface \ce{H2O} reservoir, determines the \ce{H2O} posteriors (i.e., the spectrum does not directly constrain \ce{H2O}). At the surface, \ce{H2O} is at a 50\% relative humidity, and the tropopause cold trap determines the $10^{-3}$ bar \ce{H2O} concentration. 

\subsection{Inferred surface gas fluxes} \label{sec:results_flux}

\begin{figure}[t!]
  \centering
  \includegraphics[width=\linewidth]{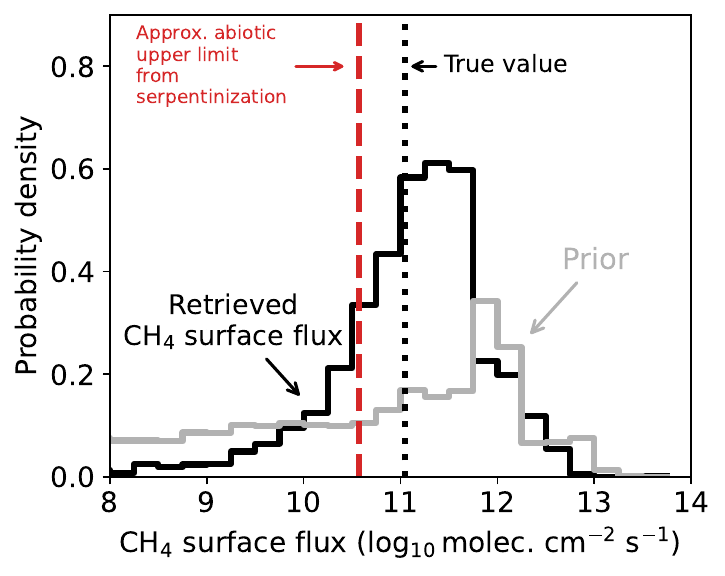}
  \caption{Inferred surface \ce{CH_4} flux for an Archean Earth-like TRAPPIST-1 e observed with 10 JWST Prism transits (Figure \ref{fig:true_atmos}). The black dotted line is the true value. The red shaded dashed line is a plausible upper limit methane flux from serpentinization (see text). The bulk of the posterior distribution resides above the region that can be explained by serpentinization, possibly favoring the presence of a biosphere.}
  \label{fig:ch4flux}
\end{figure}

We propagated the posteriors for each surface partial pressure through our forward model to derive posteriors for the surface flux of \ce{CH4}, \ce{CO2}, \ce{O2}, \ce{CO} and \ce{H2} (Appendix Figure \ref{fig:corner_flux}). The surface \ce{CH4} flux (i.e., $F_{\mathrm{CH_4}}$), shown in Figure \ref{fig:ch4flux}, is the only surface flux that is well constrained by the data. The inversion favors a $11.2_{-0.9}^{+0.6}$ $\log_{10}$ molecules cm$^{-2}$ s$^{-1}$ \ce{CH4} flux into the atmosphere (68\% credible interval) because \ce{CH4} is confidently detected by the spectra (Figures \ref{fig:true_atmos} and \ref{fig:abundances}) and \ce{CH4} is photochemically unstable. Over much of the posterior, \ce{CH4} is primarily destroyed by photolysis and reactions with O and OH. \ce{CH4} destruction often yields \ce{CH3} which is further oxidized through a sequence involving O and OH to CO and \ce{CO2} (via \ce{H2CO} and \ce{HCO}). The oxidants (O and OH) that convert \ce{CH4} to CO and \ce{CO2} are often ultimately derived from \ce{H2O} (e.g., $\ce{H2O + {\mathit{h}\nu} -> OH + H}$). Because methane oxidation is controlled by UV-driven oxidant production, we later quantify the sensitivity of the inferred $F_{\mathrm{CH_4}}$ to the assumed stellar UV spectrum (Section \ref{sec:discussion_caveat}; Figure \ref{fig:muscles}).


To determine what controls the width of the retrieved \ce{CH4} surface-flux posterior, we re-propagated the surface pressure posteriors of \ce{CH4}, \ce{CO2}, \ce{O2}, \ce{CO} and \ce{H2} through our forward model to derive a new $F_{\mathrm{CH_4}}$ probability distribution while artificially fixing one or more species to their true surface pressures. The test addresses questions like: ``if additional data perfectly measured the surface \ce{CH4} pressure, how would that improve the inferred surface \ce{CH4} flux?'' For each experiment we computed the 68\% credible-interval width of the resulting \ce{CH4} flux posterior, with results summarized in Appendix Table \ref{tab:sensitivity}. We find that uncertainty in inferred $P_\mathrm{CH_4}$ is the largest single contributor to the \ce{CH4} flux uncertainty: If the data perfectly constrained $P_\mathrm{CH_4}$, but left the other posteriors untouched the 68\% CI width would shrink by $\sim 27\%$ from 1.5 to 1.1 ($\log_{10}$ molecules cm$^{-2}$ s$^{-1}$). More decisive constraints on $F_{\mathrm{CH_4}}$ requires better knowledge of $P_\mathrm{CH_4}$, $P_\mathrm{CO}$ and $P_\mathrm{O_2}$. Knowing all three abundances shrinks the posterior width by 80\%. This result is understandable because both \ce{CO} and \ce{O2} are important for the \ce{CH4} photochemical lifetime. Large \ce{O2} can enhance \ce{CH4} destruction by delivering additional oxidants, while \ce{CO} can react with OH, inhibiting its reaction with \ce{CH4} and impacting the \ce{CH4} destruction rate.

Unfortunately, JWST is not very sensitive to either \ce{CO} or \ce{O2} in transmission \citep{KrissansenTotton2018b,LustigYaeger2019}. When repeating the Bayesian inversion, but instead for 50 transits, the 68\% width of the $F_{\mathrm{CH_4}}$ posterior shrank from 1.5 to 0.8 (i.e., by 50\%; units are $\log_{10}$ molecules cm$^{-2}$ s$^{-1}$), representing a hard limit to how precisely a JWST Prism transmission spectrum can estimate the \ce{CH4} flux.

Our inversion and sensitivity analysis also reveals that the inferred \ce{CH4} flux is not influenced by the \ce{CO2} abundance. Appendix Table \ref{tab:sensitivity} shows that more precise knowledge of $P_\mathrm{CO_2}$ does not notably tighten the $F_{\mathrm{CH_4}}$ distribution. To further test this, we repeated the posterior propagation while replacing the retrieved $P_{\mathrm{CO_2}}$ posterior with our broad prior ($\log_{10}$ uniform from $10^{-9}$ to $10^{1}$ bar), leaving the other species posteriors unchanged. This modification does not substantially alter the resulting $F_{\mathrm{CH_4}}$ distribution. Thus, within our model, uncertainty in $P_{\mathrm{CO_2}}$ is not a meaningful contributor to the uncertainty in the inferred \ce{CH4} surface flux (Figure \ref{fig:ch4flux}). This is consistent with the expectation that \ce{CH4} destruction is largely controlled by oxidants from water vapor and not \ce{CO2}.

\subsection{Probabilistic interpretation of inferred surface gas fluxes} \label{sec:results_interp}

Using the flux-based biosignature definitions established in Section \ref{sec:methods_interp}, we now calculate the probability that our inferred surface gas fluxes (Section \ref{sec:results_flux} and Appendix Figure \ref{fig:corner_flux}) are consistent with those expected from biology.

\begin{table*}[t]
\centering
\caption{Posterior and prior probability masses contained within each region of surface-flux space for the Section \ref{sec:results} inversion.}
\label{tab:flux_space_probs}
\begin{tabularx}{.96\textwidth}{lcXcc}
\hline\hline
Subset of $\mathcal{F}$ & Definition &
Description of subset &
$P(\mathbf{F}\in\mathcal{S}\mid d)^\dagger$ &
$P(\mathbf{F}\in\mathcal{S})^\ddagger$ \\
\hline
$\mathcal{H}_{\mathrm{CH}_4}$  & Eq.~\eqref{eq:def_h_ch4} & Methane flux exceeding abiotic serpentinization limit & 0.77 & 0.38 \\
$\mathcal{V}$                  & Eq.~\eqref{eq:def_v} & Reducing volcanic false positive within $\mathcal{H}_{\mathrm{CH}_4}$ & 0.06 & 0.05 \\
$\mathcal{B}_{\mathrm{CH}_4}$  & Eq.~\eqref{eq:def_b_ch4} & Archean Earth-like methane biosignature & 0.71 & 0.33 \\
$\mathcal{B}_{\mathrm{O}_2}$   & Eq.~\eqref{eq:def_b_o2} & Modern Earth-like \ce{O2} biosignature & 0.44 & 0.36 \\
$\mathcal{L}$                  & Eq.~\eqref{eq:def_l} & Union of methane and oxygen biosignature regions & 0.76 & 0.38 \\
$\mathcal{A}$                  & Eq.~\eqref{eq:def_a} & Abiotic flux region ($\mathcal{F}\setminus\mathcal{L}$) & 0.24 & 0.62 \\
\hline
\multicolumn{5}{p{0.96\textwidth}}{
  $^\dagger$Posterior probability the gas fluxes on the planet are within subset $\mathcal{S}$.

  $^\ddagger$Assumed prior probability the gas fluxes on the planet are within subset $\mathcal{S}$.
}\\
\end{tabularx}
\end{table*}

To achieve this, we compute the posterior and prior probability mass in each of the biosignature and false-positive spaces defined in Section \ref{sec:methods_interp}. In practice, we accomplish this by calculating the fraction of our posterior or prior samples in a given flux space $\mathcal{S}$. For example, for $\mathcal{S} = \mathcal{L}$ (where $\mathcal{L}$ is our biosignature definition from Section \ref{sec:methods_interp}), the posterior probability mass in $\mathcal{L}$ is,
\begin{align}
P(\mathbf{F}\in\mathcal{L}\mid d)
&=
\int_{\mathcal{L}} p(\mathbf{F}\mid d)\, d\mathbf{F} \\
&\;\approx\;
\frac{1}{N}
\sum_{i=1}^{N}
\begin{cases}
1, & \mathbf{F}_i \in \mathcal{L} \\
0, & \mathbf{F}_i \notin \mathcal{L}
\end{cases}
\end{align}
Here, $p(\mathbf{F}\mid d)$ is the posterior probability density over surface gas-flux space and should be read as ``the probability density of flux vector $\mathbf{F}$ given the spectral data $d$.'' Also, $P(\mathbf{F}\in\mathcal{L}\mid d)$ is the posterior probability that the inferred surface gas flux lies within the life-consistent region $\mathcal{L}$. 

Table \ref{tab:flux_space_probs} lists each probability mass for the prior and posterior in our nominal TRAPPIST-1 e Bayesian inversion (Section \ref{sec:results_flux}). For our synthetic test, we find a 76\% chance that the surface gas fluxes on TRAPPIST-1 e are consistent with the flux-based biosignature $\mathcal{L}$ (i.e., $P(\mathbf{F}\in\mathcal{L}\mid d) = 0.76$). For comparison, only 38\% of our adopted prior resides in the life-consistent region (i.e., $P(\mathbf{F}\in\mathcal{L}) = 0.38$), indicating that the spectrum shifts substantial probability mass toward the biosignature region relative to the prior. The fact that $P(\mathbf{F}\in\mathcal{L}\mid d)=0.76$ is only modestly larger than $P(\mathbf{F}\in\mathcal{B}_{\mathrm{CH_4}}\mid d)=0.71$ and $P(\mathbf{F}\in\mathcal{B}_{\mathrm{O_2}}\mid d)=0.44$ reflects the substantial overlap between these two subsets of flux space, so the probability of their union is much smaller than the sum of their separate probabilities.

\section{Discussion} \label{sec:discussion}

\subsection{Translating our results to a probability of life on the planet} \label{sec:discussion_prob}

Our synthetic TRAPPIST-1 e flux-inference calculation finds that 76\% of the posterior mass lies in the life-consistent flux region ($P(\mathbf{F}\in\mathcal{L}\mid d) = 0.76$). However, this quantity is not itself the probability that the planet hosts life. $P(\mathbf{F}\in\mathcal{L}\mid d)$ can be thought of as the probability of a biosignature given the spectra, and additional assumptions are necessary to translate it to a probability that a world is inhabited.

The probability we ultimately seek is $P(\mathrm{life}\mid d,c)$, the probability of life given the spectral data and planetary context, which \citet{Catling2018} derived from Bayes' theorem:
\begin{equation} \label{eq:catling}
  P(\mathrm{life}\mid d,c)
  =
  \frac{P(d \mid c, \mathrm{life})\,P(\mathrm{life}\mid c)}
  {P(d \mid c)}
\end{equation}
\begin{equation}
  \begin{aligned}
    P(d \mid c)
    &=
    P(d \mid c, \mathrm{life})\,P(\mathrm{life}\mid c) \\
    &\quad
    + P(d \mid c, \mathrm{no\ life})\,P(\mathrm{no\ life}\mid c)
  \end{aligned}
\end{equation}
Here, $d$ stands for ``spectral data'' and $c$ refers to ``planetary context'', such that $P(\mathrm{life}\mid d,c)$ should be read as ``the probability the exoplanet has life given the spectral data and planetary context.'' The term $P(\mathrm{life}\mid c)$ is a prior probability of life (discussed later), while the denominator is a normalization. We next derive an expression for $P(\mathrm{life}\mid d,c)$ in terms of quantities from our retrieval $P(\mathbf{F}\in\mathcal{L}\mid d)$ and $P(\mathbf{F}\in\mathcal{L})$. Multiplying both the right-hand-side numerator and denominator of Equation \eqref{eq:catling} by $1/P(d \mid c, \mathrm{no\ life})$ and substituting $P(\mathrm{no\ life}\mid c) = 1 - P(\mathrm{life}\mid c)$ gives
\begin{equation} \label{eq:catling_simple}
  P(\mathrm{life}\mid d,c)
  =
  \frac{K_{\mathrm{life}}\,P(\mathrm{life}\mid c)}
  {1 + (K_{\mathrm{life}} - 1)\,P(\mathrm{life}\mid c)}
\end{equation}
where $K_{\mathrm{life}}$ is the Bayes factor for the hypotheses ``life'' and ``no life'':
\begin{equation}
  K_{\mathrm{life}} = \frac{P(d \mid c, \mathrm{life})}{P(d \mid c, \mathrm{no\ life})}.
\end{equation}

We cannot compute $K_{\mathrm{life}}$ directly from our inversion outputs. Instead, we can compute the following Bayes factor:
\begin{equation} \label{eq:K_approx_main}
  \begin{aligned}
    K_{\mathcal{L}}
    &\equiv
    \frac{P(d\mid c,\mathbf{F}\in\mathcal{L})}{P(d\mid c,\mathbf{F}\in\mathcal{A})} \\
    &=
    \frac{P(\mathbf{F} \in \mathcal{L} \mid d, c)/P(\mathbf{F} \in \mathcal{A} \mid d, c)}{P(\mathbf{F} \in \mathcal{L} \mid c)/P(\mathbf{F} \in \mathcal{A} \mid c)}
  \end{aligned}
\end{equation}
where we assume that $\mathcal{A}$ is the complement of $\mathcal{L}$ in flux space (see Section \ref{sec:methods_interp}), such that $P(\mathbf{F} \in \mathcal{A} \mid d, c) = 1 - P(\mathbf{F} \in \mathcal{L} \mid d, c)$ and $P(\mathbf{F} \in \mathcal{A} \mid c) = 1 - P(\mathbf{F} \in \mathcal{L} \mid c)$. Equation \eqref{eq:K_approx_main} follows directly from Bayes' theorem in odds form. We emphasize that assuming that $\mathcal{A}$ is the complement of $\mathcal{L}$ is valid only for our simple and discrete definitions for $\mathcal{A}$ and $\mathcal{L}$. More realistically, the life-bearing and abiotic flux spaces should be treated as overlapping probability distributions, in which case Equation \eqref{eq:K_approx_main} would break down (see the discussion at the end of Section \ref{sec:methods_interp}).

We then adopt the proxy assumption $K_{\mathrm{life}} \approx K_{\mathcal{L}}$. This approximation is defensible only if our flux-based biosignature definition from Section \ref{sec:methods_interp} is a good discriminator between an inhabited and lifeless planet, and if the forward model used to infer surface fluxes captures the dominant physical and chemical processes. 
Additionally, we assume that $\mathcal{L}$ is approximately context independent, in the sense that once the sustaining surface gas fluxes are inferred, much of the dependence on planetary context that shapes atmospheric abundances, such as stellar UV and atmospheric redox state, has already been folded into the flux inference itself, so that $P(\mathbf{F} \in \mathcal{L} \mid d, c) \approx P(\mathbf{F} \in \mathcal{L} \mid d)$. In our nominal retrieval from Section \ref{sec:results}, $P(\mathbf{F} \in \mathcal{L} \mid d, c) = 0.76$ and $P(\mathbf{F} \in \mathcal{L} \mid c) = 0.38$, so in this instance, $K_{\mathcal{L}} \approx 5.1$ and, under the proxy assumption, $K_{\mathrm{life}} \approx K_{\mathcal{L}} \approx 5.1$. 

The final unknown in Equation \eqref{eq:catling_simple} is $P(\mathrm{life}\mid c)$, our prior expectation that the planet hosts life given its planetary context. This prior reflects both the likelihood that the planet is habitable (e.g., whether it resides in the habitable zone) and the probability that life emerges and persists over geologic time. At present, $P(\mathrm{life}\mid c)$ is poorly constrained, as it represents a population-level statistic about the prevalence of life that has not yet been empirically measured. 

For example, an optimistic prior of $P(\mathrm{life}\mid c) = 0.5$ would yield $P(\mathrm{life}\mid d,c) = 0.84$ for our nominal case, whereas a more pessimistic prior of $P(\mathrm{life}\mid c)=10^{-3}$ would imply $P(\mathrm{life}\mid d,c) = 0.005$. However, this prior dependence does not imply that life detection is fundamentally hopeless: if the Bayes factor $K_{\mathrm{life}}$ becomes sufficiently large (i.e., if $P(d\mid c,\mathrm{no\ life}) \rightarrow 0$), then $P(\mathrm{life}\mid d,c)$ approaches unity regardless of the assumed prior. The role of the prior is therefore to modulate the strength of the evidence rather than to negate it, underscoring that progress in exoplanet life detection will come both from stronger spectral constraints and from improved understanding of life's origin and prevalence \citep{Wong2022}.

\begin{figure*}[t!]
  \centering
  \includegraphics[width=\textwidth]{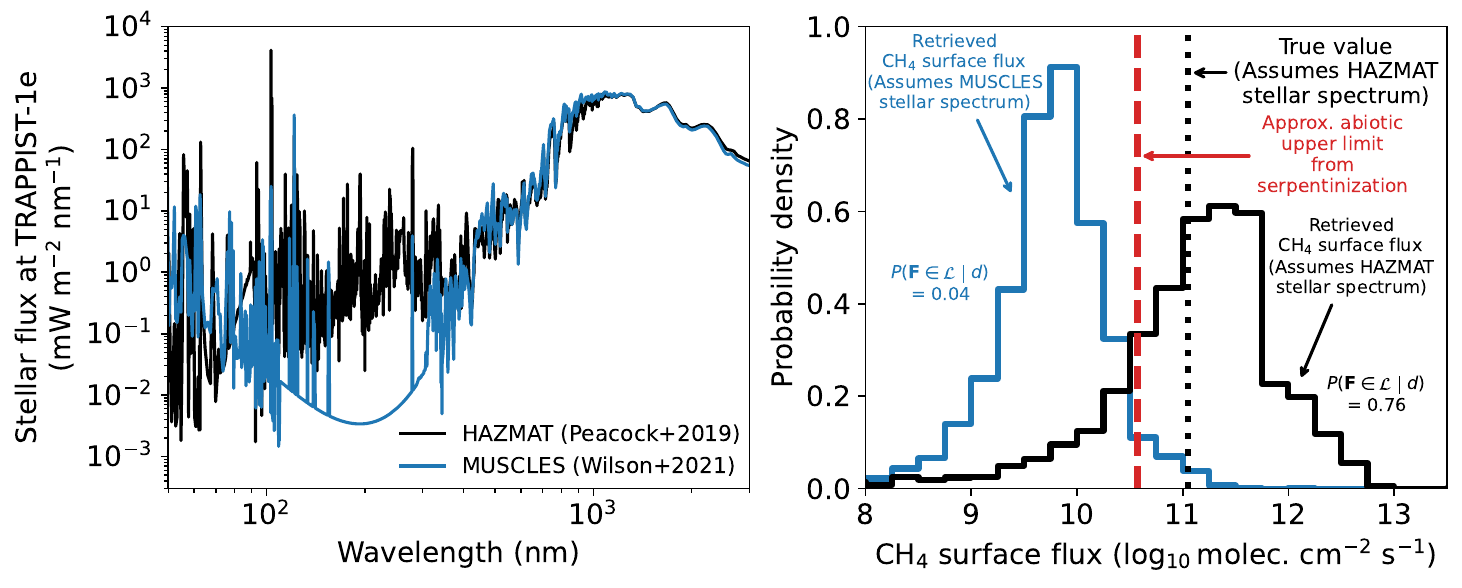}
  \caption{The sensitivity of the inferred surface \ce{CH4} flux to TRAPPIST-1's stellar spectrum. Left: The HAZMAT TRAPPIST-1 spectrum \citep[black line;][]{Peacock2019} that we nominally assume in this work alongside the MUSCLES TRAPPIST-1 semi-empirical spectrum \citep[blue line;][]{Wilson2021} that we use in this sensitivity test. Right: The inferred surface methane flux for the Figure \ref{fig:true_atmos} synthetic JWST data (10 transits) assuming the HAZMAT TRAPPIST-1 spectrum (black line) and MUSCLES TRAPPIST-1 spectrum (blue line). The retrieval that uses the MUSCLES spectrum can explain the \ce{CH4} spectral features with a small \ce{CH4} flux, demonstrating that stellar UV needs to be better constrained to interpret methane biosignatures on TRAPPIST-1 e.}
  \label{fig:muscles}
\end{figure*}

\subsection{Caveat: Our inversion for surface fluxes does not marginalize over all uncertainty} \label{sec:discussion_caveat}

The TRAPPIST-1 e flux-inference calculation described here fixes several variables that are not well constrained. For example, we only consider atmospheres with 1 bar of \ce{N2}, a surface liquid water reservoir (i.e., an ocean), a constant vertical mixing coefficient, and irradiated by a single plausible stellar UV spectrum. Furthermore, our inversion does not marginalize over the uncertainty in the chemical kinetics or opacities themselves. A better calculation would treat these unknowns as free parameters. However, to accomplish this, photochemical and climate models will need to be more efficient, as currently the calculations shown here, which do not exhaust the parameter space, expend significant computational resources (e.g., the Section \ref{sec:method_grid} pre-computed grid requires $\sim 4000$ core-hours).

Here, we demonstrate the impact of one of these uncertain variables: the incident stellar UV. This sensitivity test is motivated by \citet{sneedprep}, who showed that varying the assumed TRAPPIST-1 UV spectrum can substantially alter atmospheric outcomes for photochemical simulations with fixed methane fluxes. Our exercise is different: rather than prescribing the methane flux, we infer the surface gas fluxes required to sustain atmospheres consistent with the synthetic spectrum under an assumed stellar UV spectrum.

TRAPPIST-1 e's UV spectrum is still uncertain because the star is faint in the UV, large parts of the UV-EUV range cannot be observed directly, and the UV emission is variable \citep[e.g.,][]{Wilson2021}. In our nominal calculations in Section \ref{sec:results}, we adopted a TRAPPIST-1 e UV spectrum from \citet{Peacock2019}, based on \texttt{PHOENIX} upper-atmosphere stellar models. Using new HST spectroscopy and XMM-Newton X-ray data, \citet{Wilson2021} shows that the \citet{Peacock2019} models reproduced TRAPPIST-1's $\mathrm{Ly}\alpha$ and some other strong lines, but over-predicted several other UV emission lines, so they constructed a new semi-empirical spectrum for atmospheric modeling (blue line in Figure \ref{fig:muscles}, left). To quantify how uncertainty in the stellar UV propagates into uncertainty in the \ce{CH4} flux, we repeat the surface flux inference for the Figure \ref{fig:true_atmos} data using the newer \citet{Wilson2021} spectrum.

The blue distribution in Figure \ref{fig:muscles}, right, shows that the derived \ce{CH4} flux when assuming the \citet{Wilson2021} UV spectrum is one or two orders of magnitude smaller on average than the inversion that uses the \cite{Peacock2019} spectrum. The difference is almost completely accounted for by the discrepant $130$ to $230$ nm continuum between the two spectra. We demonstrate this by running two models for our ``true'' Archean atmospheric composition (Figure \ref{fig:true_atmos}). The first uses the \citet{Wilson2021} spectrum, yielding a \ce{CH4} flux a factor of $\sim 20$ smaller than the same calculation with the \cite{Peacock2019} spectrum. We also performed a second calculation with the \citet{Wilson2021} spectrum except with the 130 to 230 nm stellar flux replaced by the \cite{Peacock2019} spectrum. The resulting \ce{CH4} flux matches the nominal \cite{Peacock2019} simulation (Figure \ref{fig:true_atmos}) to within $\sim5\%$. The low 130 to 230 nm stellar flux in the \citet{Wilson2021} spectrum leads to a slow $1.9\times10^{9}$ molecules cm$^{-2}$ s$^{-1}$ \ce{H2O} photolysis rate (compared to $2.1 \times 10^{11}$ molecules cm$^{-2}$ s$^{-1}$ in the \cite{Peacock2019} model), inhibiting OH and O production and methane oxidation to CO and \ce{CO2}. These results demonstrate that stellar UV emission must be better constrained to quantitatively interpret methane biosignatures on TRAPPIST-1 e.

This example highlights the importance of marginalizing over all unknown variables when inferring surface gas fluxes. Without a full propagation of the uncertainties, the biosignature and probabilistic framework developed here (Sections \ref{sec:methods_interp}, \ref{sec:results_interp} and \ref{sec:discussion_prob}) should be interpreted with caution and understood as conditional on the modeling assumptions adopted in the inversion.






\section{Conclusions}

Many spectral retrieval models infer surface atmospheric abundances, but abundances are challenging to interpret as a sign of life given their indirect relationship to metabolism on a planetary surface. Here, we developed a method to instead infer surface gas fluxes from telescope spectra by inverting a coupled photochemical-climate model (\texttt{Photochem}) together with a radiative transfer code (\texttt{PICASO}). We applied the method to a synthetic JWST NIRSpec Prism transmission spectrum of TRAPPIST-1 e, assuming the planet is inhabited by an Archean-Earth like biosphere. Finally, we show how surface gas fluxes can be readily and probabilistically interpreted as a sign of life. From this analysis, our conclusions are the following:

\begin{itemize}
\item The inversion, which embeds a photochemical model, enables constraints on gases that do not substantially impact the observed spectra but are photochemically tied to other species that have large spectral features. In our TRAPPIST-1 e example, although these species are observationally unconstrained, we nevertheless infer peaked posteriors for stratospheric \ce{O2}, \ce{H2} and \ce{CO}, because these molecules are photochemical byproducts of the confidently detected \ce{CH4} and \ce{CO2}.
\item In our nominal Archean TRAPPIST-1 e test case, \ce{CH4} is the only gas with a well-constrained surface gas flux, with a retrieved value of $\log_{10}F_{\mathrm{CH_4}} = 11.2^{+0.6}_{-0.9}$ molecules cm$^{-2}$ s$^{-1}$ (68\% credible interval), demonstrating that JWST can perhaps, in specific scenarios, meaningfully constrain biosphere-relevant boundary conditions.
\item The width of the \ce{CH4} surface flux ($F_{\mathrm{CH_4}}$) posterior is driven primarily by uncertainty in the methane abundance itself and secondarily by the CO and \ce{O2} abundances, because CO and \ce{O2} modulate oxidant budgets (e.g., O and OH) and therefore methane lifetime. On the other hand, the \ce{CO2} abundance does not contribute to the uncertainty in $F_{\mathrm{CH_4}}$.
\item We show how biosignatures and false-positives can be cleanly expressed in terms of surface gas fluxes. For example, we define an Archean Earth-like biosignature as the region of flux space where $F_{\mathrm{CH_4}}$ is larger than what is plausible from ocean-bottom serpentinization and where the CO and \ce{H2} fluxes are inconsistent with a large abiotic source of \ce{CH4} from reducing volcanism.
\item For our nominal retrieval, the posterior probability mass in the life-consistent flux region is two times larger than the prior: $P(\mathbf{F}\in \mathcal{L} \mid d) \approx 0.76$ vs.\ $P(\mathbf{F}\in \mathcal{L}) \approx 0.38$.
\item The probability we ultimately seek, $P(\mathrm{life}\mid d,c)$, depends on the prior probability of life in a given planetary context ($P(\mathrm{life}\mid c)$). While this prior is uncertain, sufficiently decisive spectral evidence (e.g., $P(\mathbf{F}\in \mathcal{L} \mid d) \rightarrow 1$) can dominate even pessimistic assumptions for $P(\mathrm{life}\mid c)$.
\item Our nominal TRAPPIST-1 e test case does not marginalize over all uncertainties. For example, adopting an alternative TRAPPIST-1 UV spectrum shifts the inferred methane flux by orders of magnitude, resulting in a very low probability that life impacts the planetary environment, even though a biosphere is present. This underscores that a robust search for life with JWST on TRAPPIST-1 e will require better UV constraints and marginalization over other model uncertainties.

\end{itemize}

In sum, the search for life on exoplanets requires more than the detection of putative biogenic gases. It requires evidence that gases in a planetary atmosphere are sustained by surface gas fluxes consistent with metabolism, and inconsistent with known abiotic processes. Here, we introduce a framework that enables this inference by using atmospheric chemistry and physics to link spectra to biological and abiotic processes at a planet's surface.

\begin{acknowledgments}

We thank our reviewer who improved the quality of this article. We also thank Mark Moussa and Giada Arney for conversations that aided this research. This work was supported in part by the NASA Planetary Science Division ``Center for Life Detection'' ISFM, and NASA's Interdisciplinary Consortia for Astrobiology Research (grant No. NNH19ZDA001N-ICAR) under award number 19-ICAR19\_2-0041. Also, NFW was supported in part by the NASA Postdoctoral Program. JKT and VSM were supported by a NASA Astrophysics Decadal Survey Precursor Science grant 80NSSC23K1471 and the Virtual Planetary Laboratory, a member of the NASA Nexus for Exoplanet System Science (NExSS), funded via NASA Astrobiology Program grant No. 80NSSC23K1398.  JKT was also supported by the Alfred P. Sloan Foundation under grant No. 2025-2520. AVY would like to acknowledge support from the GSFC Sellers Exoplanet Environments Collaboration (SEEC), which is supported by NASA's Planetary Science Division's Research Program. ELS and EWS acknowledge support from the NASA Interdisciplinary Consortia for Astrobiology Research (ICAR) Program via the Alternative Earths team, with funding issued under grant No. 80NSSC21K0594.

\end{acknowledgments}

\appendix
\setcounter{figure}{0}
\setcounter{table}{0}
\renewcommand{\thefigure}{A\arabic{figure}}
\renewcommand{\thetable}{A\arabic{table}}

\section{Supplemental Figures and Tables}

\begin{figure*}
  \centering
  \includegraphics[width=\textwidth]{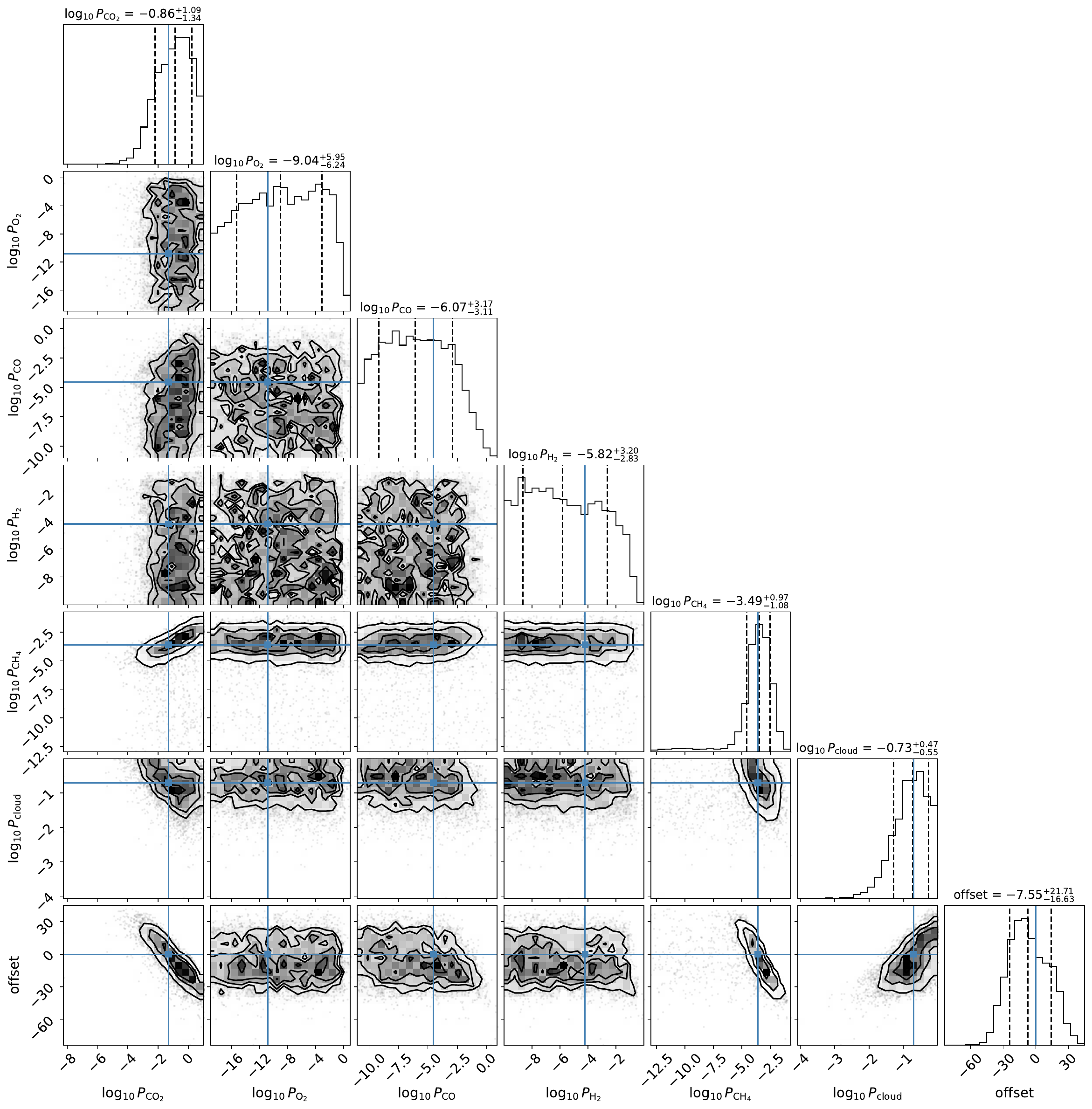}
  \caption{The corner plot for our nominal retrieval of the synthetic spectrum in Figure \ref{fig:true_atmos} (10 JWST Prism transits). The parameters $\log_{10}P_\mathrm{CO_2}$, $\log_{10}P_\mathrm{O_2}$, $\log_{10}P_\mathrm{CO}$, $\log_{10}P_\mathrm{H_2}$, $\log_{10}P_\mathrm{CH_4}$ are surface partial pressures in $\log_{10}$ bars. $\log_{10}P_\mathrm{cloud}$ is the gray cloud-top pressure in $\log_{10}$ bars. The ``offset'' parameter is an offset added to the spectrum in ppm. The blue lines indicate the true values.}
  \label{fig:corner}
\end{figure*}

\begin{figure*}
  \centering
  \includegraphics[width=\textwidth]{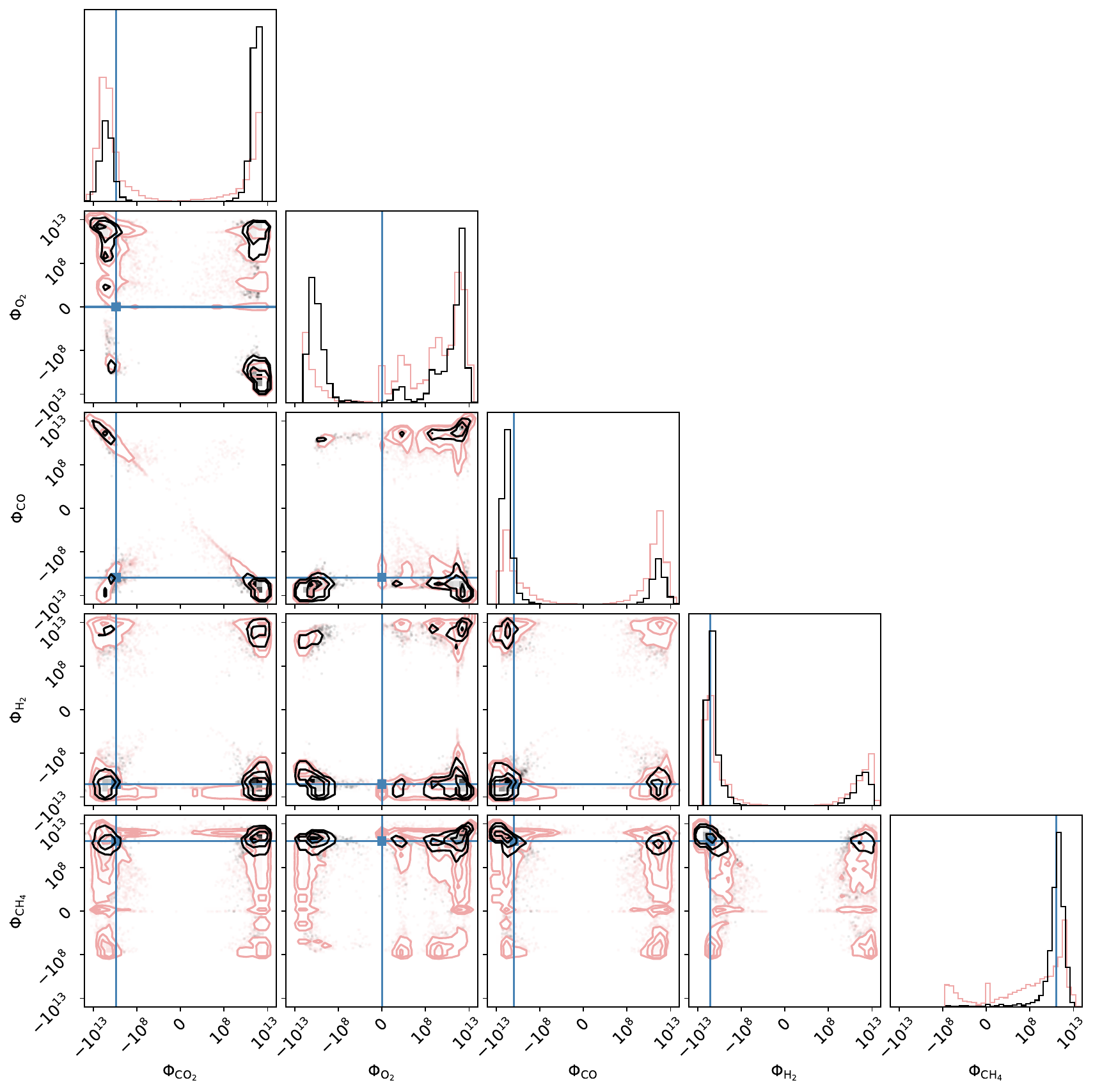}
  \caption{The corner plot for surface gas fluxes in our nominal inversion of the synthetic spectrum in Figure \ref{fig:true_atmos}. Each parameter is a surface gas flux in $\log_{10}$ molecules cm$^{-2}$ s$^{-1}$. Positive values are fluxes into the atmosphere while negative values are fluxes into the planet. Black distributions are posteriors, red distributions are priors, and the blue lines are true values. Note that most posteriors appear bimodal at first glance, but this is only an artifact of the semi-log scale (necessary for visual clarity).}
  \label{fig:corner_flux}
\end{figure*}

This appendix collects supporting diagnostics referenced in the main text. Table \ref{tab:sensitivity} quantifies how fixing selected retrieved abundances to their true values affects the inferred \ce{CH4} flux posterior width. Figure \ref{fig:corner} and Figure \ref{fig:corner_flux} provide the full posterior structure for atmospheric abundances and surface fluxes, respectively.

\begin{table*}
\centering
\caption{The sensitivity of the surface \ce{CH4} flux posterior on abundance constraints.} \label{tab:sensitivity}
\setlength{\tabcolsep}{3pt} 

\begin{tabular*}{0.84\textwidth}{@{\extracolsep{\fill}}l|*{21}{c}@{}}
\hline
\hline
\multicolumn{1}{c|}{Species} &
\multicolumn{21}{c}{Each column is one configuration of species fixed to their true value (\checkmark\ = fixed, $\cdot$ = not fixed)} \\
\hline
CH$_4$ &
$\cdot$ & $\checkmark$ & $\cdot$ & $\cdot$ & $\cdot$ & $\cdot$ &
$\checkmark$ & $\checkmark$ & $\checkmark$ & $\checkmark$ &
$\checkmark$ & $\checkmark$ & $\checkmark$ & $\checkmark$ &
$\checkmark$ & $\checkmark$ & $\checkmark$ & $\checkmark$ &
$\checkmark$ & $\checkmark$ & $\cdot$ \\
CO$_2$ &
$\cdot$ & $\cdot$ & $\checkmark$ & $\cdot$ & $\cdot$ & $\cdot$ &
$\checkmark$ & $\cdot$ & $\cdot$ & $\cdot$ &
$\checkmark$ & $\checkmark$ & $\checkmark$ & $\cdot$ &
$\cdot$ & $\cdot$ & $\checkmark$ & $\checkmark$ &
$\checkmark$ & $\cdot$ & $\checkmark$ \\
O$_2$ &
$\cdot$ & $\cdot$ & $\cdot$ & $\checkmark$ & $\cdot$ & $\cdot$ &
$\cdot$ & $\checkmark$ & $\cdot$ & $\cdot$ &
$\checkmark$ & $\cdot$ & $\cdot$ & $\checkmark$ &
$\checkmark$ & $\cdot$ & $\checkmark$ & $\checkmark$ &
$\cdot$ & $\checkmark$ & $\checkmark$ \\
CO &
$\cdot$ & $\cdot$ & $\cdot$ & $\cdot$ & $\checkmark$ & $\cdot$ &
$\cdot$ & $\cdot$ & $\checkmark$ & $\cdot$ &
$\cdot$ & $\checkmark$ & $\cdot$ & $\checkmark$ &
$\cdot$ & $\checkmark$ & $\checkmark$ & $\cdot$ &
$\checkmark$ & $\checkmark$ & $\checkmark$ \\
H$_2$ &
$\cdot$ & $\cdot$ & $\cdot$ & $\cdot$ & $\cdot$ & $\checkmark$ &
$\cdot$ & $\cdot$ & $\cdot$ & $\checkmark$ &
$\cdot$ & $\cdot$ & $\checkmark$ & $\cdot$ &
$\checkmark$ & $\checkmark$ & $\cdot$ & $\checkmark$ &
$\checkmark$ & $\checkmark$ & $\checkmark$ \\
\hline
CI width &
1.5 & 1.1 & 1.6 & 1.2 & 1.4 & 1.4 &
1.0 & 0.9 & 0.6 & 1.1 &
0.8 & 0.6 & 1.1 & 0.3 &
0.7 & 0.6 & 0.1 & 0.7 &
0.6 & 0.2 & 1.3 \\
\hline
\multicolumn{22}{@{}p{0.84\textwidth}@{}}{Notes: ``CI width'' is the 68\% credible interval width of $\log_{10}F_{\mathrm{CH_4}}$ in $\log_{10}$ molecules cm$^{-2}$ s$^{-1}$. The first column, where no species is fixed, gives the $\log_{10}F_{\mathrm{CH_4}}$ posterior width shown in Figure \ref{fig:ch4flux}. Each other column fixes one or more gas to their true surface pressure, then propagates those abundances through the forward model to derive a new \ce{CH4} flux distribution, with a given 68\% CI width.}
\end{tabular*}
\end{table*}

\section{Justification for loosely coupling the photochemical and climate models} \label{sec:appendix_couple}

As described in Section \ref{sec:method_pc}, we compute atmospheric states by ``loosely`` coupling the photochemical and climate models: the climate model is first run for a given set of input surface \ce{CO2}, \ce{O2}, \ce{CO}, \ce{H2} and \ce{CH4} partial pressures, then, with the output P-T profile, the photochemistry is computed. No further iterations are performed between each code to ensure a fully self-consistent state. We chose this approximation to make our pre-computed grid (Section \ref{sec:method_grid}) computationally feasible. Here, we illustrate the error introduced by this loose-coupling approach near our assumed ``true'' Archean Earth-like atmosphere (Section \ref{sec:method_case}).

\begin{figure*}
  \centering
  \includegraphics[width=.5\textwidth]{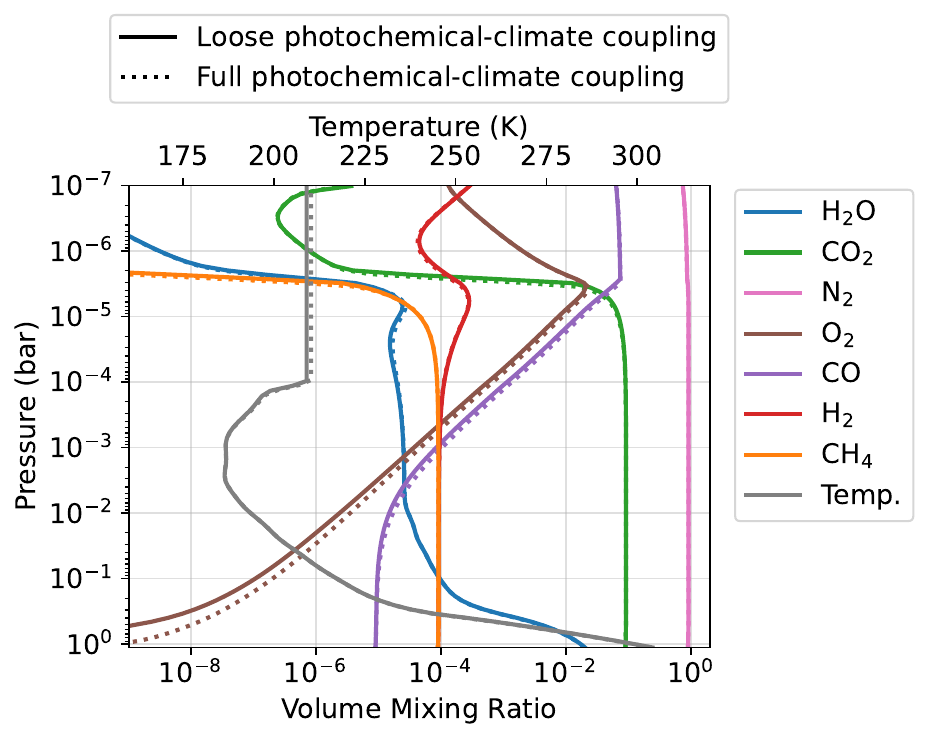}
  \caption{Two photochemical-climate simulations of TRAPPIST-1 e with an Archean Earth-like atmosphere. The solid lines are a simulation that adopts the ``loose'' coupling strategy that we use throughout this article. The dotted lines are a fully self-consistent photochemical-climate state. The composition and pressure profiles are very similar for both methods, justifying our ``loose'' coupling approach.}
  \label{fig:couple}
\end{figure*}

Figure \ref{fig:couple} shows two atmospheric simulations at the following input partial pressures (bars): $P_\mathrm{CO_2} = 10^{1}$, $P_\mathrm{O_2} = 10^{-11}$, $P_\mathrm{CO} = 10^{-5}$, $P_\mathrm{H_2} = 10^{-4}$ and $P_\mathrm{CH_4} = 10^{-4}$. As with all other calculations in this paper, the simulation also has 1 bar of \ce{N2} and a surface \ce{H2O} reservoir. This is one of the closest points in our grid (Section \ref{sec:method_grid}) to the ``true'' Archean Earth-like atmosphere that we assume in our flux-inference case study (Section \ref{sec:method_case}). The solid lines are from a simulation that uses our ``loose'' photochemical-climate coupling approach (adopted throughout this paper), while the dotted lines are for a fully self-consistent photochemical-climate calculation. Both models have very similar composition and temperature profiles (and have surface \ce{CH4} fluxes within 25\%), because the photochemistry does not significantly alter the composition of the main climate-relevant absorbers (\ce{H2O}, \ce{CO2} and \ce{CH4}). However, we find that even \ce{O2}-rich test cases with substantial photochemical \ce{O3} (a strong stratospheric absorber) are well approximated by our loose coupling strategy, which neglects \ce{O3} in the radiative-convective calculation. A likely reason is that TRAPPIST-1 emits predominantly in the near-infrared, where \ce{H2O} already strongly absorbs, reducing the relative radiative importance of stratospheric \ce{O3}.

In summary, the loose coupling approach adopted throughout this paper is a sensible approximation, especially near our assumed ``true'' Archean Earth-like atmosphere.

\section{Derivation of maximum upward flux of \ce{CO} and \ce{H2}} \label{sec:appendix_deriv1}

A net surface flux can be decomposed into upward and downward components,
\begin{equation}
  F_i = F_i^\uparrow - F_i^\downarrow,
\end{equation}
where $F_i^\uparrow \ge 0$ and $F_i^\downarrow \ge 0$. Following \citet{Kharecha2005}, we approximate an upper bound on the downward flux of \ce{H2} and \ce{CO} with
\begin{equation}
  F_{i,\max}^\downarrow = n_i v_{i,\max},
\end{equation}
for $v_{\mathrm{CO,max}} = 1.2\times10^{-4}$ cm s$^{-1}$ and $v_{\mathrm{H_2,max}} = 2.4\times10^{-4}$ cm s$^{-1}$. Here, $n_i$ is the near surface number density (molecules cm$^{-3}$) of species $i$. It then follows that
\begin{equation}
  F_i^\uparrow = F_i + F_i^\downarrow \le F_i + n_i v_{i,\max} \equiv F_{i,\max}^\uparrow.
\end{equation}
In practice, we also enforce $F_i^\uparrow \ge 0$ by setting
\begin{equation}
  F_{i,\max}^\uparrow = \max\!\left(0,\,F_i + n_i v_{i,\max}\right)
\end{equation}

\bibliography{bib}

@article{Catling2018,
author = {Catling, David C. and Krissansen-Totton, Joshua and Kiang, Nancy Y. and Crisp, David and Robinson, Tyler D. and DasSarma, Shiladitya and Rushby, Andrew J. and Del Genio, Anthony and Bains William and Shawn, Domagal-Goldman},
   title = {Exoplanet biosignatures: A framework for their assessment},
   journal = {Astrobiol.},
   volume = {18},
   pages = {709-738},
   DOI = {10.1089/ast.2017.1737},
   url = {https://www.liebertpub.com/doi/abs/10.1089/ast.2017.1737},
   year = {2018},
   type = {Journal Article}
}

@book{Catling2017,
  title={Atmospheric evolution on inhabited and lifeless worlds},
  author={Catling, David C and Kasting, James F},
  year={2017},
  publisher={Cambridge University Press}
}

@book{Seinfeld2016,
  title={Atmospheric chemistry and physics: from air pollution to climate change},
  author={Seinfeld, John H and Pandis, Spyros N},
  year={2016},
  publisher={John Wiley \& Sons}
}

@ARTICLE{Thompson2022,
       author = {{Thompson}, Maggie A. and {Krissansen-Totton}, Joshua and {Wogan}, Nicholas and {Telus}, Myriam and {Fortney}, Jonathan J.},
        title = "{The case and context for atmospheric methane as an exoplanet biosignature}",
      journal = {Proceedings of the National Academy of Science},
         year = 2022,
        month = apr,
       volume = {119},
       number = {14},
          eid = {e2117933119},
        pages = {e2117933119},
          doi = {10.1073/pnas.2117933119},
archivePrefix = {arXiv},
       eprint = {2204.04257},
 primaryClass = {astro-ph.EP},
       adsurl = {https://ui.adsabs.harvard.edu/abs/2022PNAS..11917933T}
}

@ARTICLE{Toon1989,
       author = {{Toon}, Owen B. and {McKay}, C.~P. and {Ackerman}, T.~P. and {Santhanam}, K.},
        title = "{Rapid calculation of radiative heating rates and photodissociation rates in inhomogeneous multiple scattering atmospheres}",
      journal = {\jgr},
         year = 1989,
        month = nov,
       volume = {94},
        pages = {16287-16301},
          doi = {10.1029/JD094iD13p16287},
       adsurl = {https://ui.adsabs.harvard.edu/abs/1989JGR....9416287T}
}

@ARTICLE{Amundsen2017,
       author = {{Amundsen}, David S. and {Tremblin}, Pascal and {Manners}, James and {Baraffe}, Isabelle and {Mayne}, Nathan J.},
        title = "{Treatment of overlapping gaseous absorption with the correlated-k method in hot Jupiter and brown dwarf atmosphere models}",
      journal = {\aap},
         year = 2017,
        month = feb,
       volume = {598},
          eid = {A97},
        pages = {A97},
          doi = {10.1051/0004-6361/201629322},
archivePrefix = {arXiv},
       eprint = {1610.01389},
 primaryClass = {astro-ph.EP},
       adsurl = {https://ui.adsabs.harvard.edu/abs/2017A&A...598A..97A}
}

@ARTICLE{Graham2021,
       author = {{Graham}, R.~J. and {Lichtenberg}, Tim and {Boukrouche}, Ryan and {Pierrehumbert}, Raymond T.},
        title = "{A Multispecies Pseudoadiabat for Simulating Condensable-rich Exoplanet Atmospheres}",
      journal = {\psj},
         year = 2021,
        month = oct,
       volume = {2},
       number = {5},
          eid = {207},
        pages = {207},
          doi = {10.3847/PSJ/ac214c},
archivePrefix = {arXiv},
       eprint = {2108.12902},
 primaryClass = {astro-ph.EP},
       adsurl = {https://ui.adsabs.harvard.edu/abs/2021PSJ.....2..207G}
}

@ARTICLE{Molliere2019,
       author = {{Molli{\`e}re}, P. and {Wardenier}, J.~P. and {van Boekel}, R. and {Henning}, Th. and {Molaverdikhani}, K. and {Snellen}, I.~A.~G.},
        title = "{petitRADTRANS. A Python radiative transfer package for exoplanet characterization and retrieval}",
      journal = {\aap},
         year = 2019,
        month = jul,
       volume = {627},
          eid = {A67},
        pages = {A67},
          doi = {10.1051/0004-6361/201935470},
archivePrefix = {arXiv},
       eprint = {1904.11504},
 primaryClass = {astro-ph.EP},
       adsurl = {https://ui.adsabs.harvard.edu/abs/2019A&A...627A..67M}
}

@misc{Photochem2025,
  author       = {Nicholas Wogan},
  title        = {Photochem},
  month        = jul,
  year         = 2025,
  publisher    = {Zenodo},
  version      = {v0.6.7},
  doi          = {10.5281/zenodo.16322107},
  url          = {https://doi.org/10.5281/zenodo.16322107},
}

@ARTICLE{Kozakis2022,
       author = {{Kozakis}, Thea and {Mendon{\c{c}}a}, Jo{\~a}o M. and {Buchhave}, Lars A.},
        title = "{Is ozone a reliable proxy for molecular oxygen?. I. The O$_{2}$-O$_{3}$ relationship for Earth-like atmospheres}",
      journal = {\aap},
         year = 2022,
        month = sep,
       volume = {665},
          eid = {A156},
        pages = {A156},
          doi = {10.1051/0004-6361/202244164},
archivePrefix = {arXiv},
       eprint = {2208.09415},
 primaryClass = {astro-ph.EP},
       adsurl = {https://ui.adsabs.harvard.edu/abs/2022A&A...665A.156K}
}

@ARTICLE{Segura2005,
       author = {{Segura}, Ant{\'\i}gona and {Kasting}, James F. and {Meadows}, Victoria and {Cohen}, Martin and {Scalo}, John and {Crisp}, David and {Butler}, Rebecca A.~H. and {Tinetti}, Giovanna},
        title = "{Biosignatures from Earth-Like Planets Around M Dwarfs}",
      journal = {Astrobiology},
         year = 2005,
        month = dec,
       volume = {5},
       number = {6},
        pages = {706-725},
          doi = {10.1089/ast.2005.5.706},
archivePrefix = {arXiv},
       eprint = {astro-ph/0510224},
 primaryClass = {astro-ph},
       adsurl = {https://ui.adsabs.harvard.edu/abs/2005AsBio...5..706S}
}

@ARTICLE{MacDonald2017,
       author = {{MacDonald}, Ryan J. and {Madhusudhan}, Nikku},
        title = "{HD 209458b in new light: evidence of nitrogen chemistry, patchy clouds and sub-solar water}",
      journal = {\mnras},
         year = 2017,
        month = aug,
       volume = {469},
       number = {2},
        pages = {1979-1996},
          doi = {10.1093/mnras/stx804},
archivePrefix = {arXiv},
       eprint = {1701.01113},
 primaryClass = {astro-ph.EP},
       adsurl = {https://ui.adsabs.harvard.edu/abs/2017MNRAS.469.1979M}
}

@ARTICLE{Mukherjee2021,
       author = {{Mukherjee}, Sagnick and {Batalha}, Natasha E. and {Marley}, Mark S.},
        title = "{Cloud Parameterizations and their Effect on Retrievals of Exoplanet Reflection Spectroscopy}",
      journal = {\apj},
         year = 2021,
        month = apr,
       volume = {910},
       number = {2},
          eid = {158},
        pages = {158},
          doi = {10.3847/1538-4357/abe53b},
archivePrefix = {arXiv},
       eprint = {2102.05305},
 primaryClass = {astro-ph.EP},
       adsurl = {https://ui.adsabs.harvard.edu/abs/2021ApJ...910..158M}
}

@ARTICLE{LopezMorales2019,
       author = {{L{\'o}pez-Morales}, Mercedes and {Ben-Ami}, Sagi and {Gonzalez-Abad}, Gonzalo and {Garc{\'\i}a-Mej{\'\i}a}, Juliana and {Dietrich}, Jamie and {Szentgyorgyi}, Andrew},
        title = "{Optimizing Ground-based Observations of O$_{2}$ in Earth Analogs}",
      journal = {\aj},
         year = 2019,
        month = jul,
       volume = {158},
       number = {1},
          eid = {24},
        pages = {24},
          doi = {10.3847/1538-3881/ab21d7},
archivePrefix = {arXiv},
       eprint = {1905.05862},
 primaryClass = {astro-ph.EP},
       adsurl = {https://ui.adsabs.harvard.edu/abs/2019AJ....158...24L}
}

@ARTICLE{Meadows2023,
       author = {{Meadows}, Victoria S. and {Lincowski}, Andrew P. and {Lustig-Yaeger}, Jacob},
        title = "{The Feasibility of Detecting Biosignatures in the TRAPPIST-1 Planetary System with JWST}",
      journal = {\psj},
         year = 2023,
        month = oct,
       volume = {4},
       number = {10},
          eid = {192},
        pages = {192},
          doi = {10.3847/PSJ/acf488},
       adsurl = {https://ui.adsabs.harvard.edu/abs/2023PSJ.....4..192M}
}

@ARTICLE{KrissansenTotton2025,
       author = {{Krissansen-Totton}, Joshua and {Ulses}, Anna Grace and {Frissell}, Maxwell and {Gilbert-Janizek}, Samantha and {Young}, Amber and {Lustig-Yaeger}, Jacob and {Robinson}, Tyler and {Olson}, Stephanie and {Alei}, Eleonora and {Arney}, Giada and {Hagee}, Celeste and {Harman}, Chester and {Hinkel}, Natalie and {Lafleche}, Emilie and {Latouf}, Natasha and {Mandell}, Avi and {Moussa}, Mark M. and {Parenteau}, Niki and {Ranjan}, Sukrit and {Russell}, Blair and {Schwieterman}, Edward W. and {Sousa-Silva}, Clara and {Tokadjian}, Armen and {Wogan}, Nicholas},
        title = "{Wavelength Requirements for Life Detection via Reflected Light Spectroscopy of Rocky Exoplanets}",
      journal = {arXiv e-prints},
         year = 2025,
        month = jul,
          eid = {arXiv:2507.14771},
        pages = {arXiv:2507.14771},
          doi = {10.48550/arXiv.2507.14771},
archivePrefix = {arXiv},
       eprint = {2507.14771},
 primaryClass = {astro-ph.EP},
       adsurl = {https://ui.adsabs.harvard.edu/abs/2025arXiv250714771K}
}

@ARTICLE{KrissansenTotton2018a,
       author = {{Krissansen-Totton}, Joshua and {Olson}, Stephanie and {Catling}, David C.},
        title = "{Disequilibrium biosignatures over Earth history and implications for detecting exoplanet life}",
      journal = {Science Advances},
         year = 2018,
        month = jan,
       volume = {4},
       number = {1},
        pages = {eaao5747},
          doi = {10.1126/sciadv.aao5747},
archivePrefix = {arXiv},
       eprint = {1801.08211},
 primaryClass = {astro-ph.EP},
       adsurl = {https://ui.adsabs.harvard.edu/abs/2018SciA....4.5747K}
}

@ARTICLE{KrissansenTotton2018b,
       author = {{Krissansen-Totton}, Joshua and {Garland}, Ryan and {Irwin}, Patrick and {Catling}, David C.},
        title = "{Detectability of Biosignatures in Anoxic Atmospheres with the James Webb Space Telescope: A TRAPPIST-1e Case Study}",
      journal = {\aj},
         year = 2018,
        month = sep,
       volume = {156},
       number = {3},
          eid = {114},
        pages = {114},
          doi = {10.3847/1538-3881/aad564},
archivePrefix = {arXiv},
       eprint = {1808.08377},
 primaryClass = {astro-ph.EP},
       adsurl = {https://ui.adsabs.harvard.edu/abs/2018AJ....156..114K}
}

@ARTICLE{Wogan2020,
       author = {{Wogan}, Nicholas and {Krissansen-Totton}, Joshua and {Catling}, David C.},
        title = "{Abundant Atmospheric Methane from Volcanism on Terrestrial Planets Is Unlikely and Strengthens the Case for Methane as a Biosignature}",
      journal = {\psj},
         year = 2020,
        month = dec,
       volume = {1},
       number = {3},
          eid = {58},
        pages = {58},
          doi = {10.3847/PSJ/abb99e},
archivePrefix = {arXiv},
       eprint = {2009.07761},
 primaryClass = {astro-ph.EP},
       adsurl = {https://ui.adsabs.harvard.edu/abs/2020PSJ.....1...58W}
}

@ARTICLE{Peacock2019,
       author = {{Peacock}, Sarah and {Barman}, Travis and {Shkolnik}, Evgenya L. and {Hauschildt}, Peter H. and {Baron}, E.},
        title = "{Predicting the Extreme Ultraviolet Radiation Environment of Exoplanets around Low-mass Stars: The TRAPPIST-1 System}",
      journal = {\apj},
         year = 2019,
        month = feb,
       volume = {871},
       number = {2},
          eid = {235},
        pages = {235},
          doi = {10.3847/1538-4357/aaf891},
archivePrefix = {arXiv},
       eprint = {1812.06159},
 primaryClass = {astro-ph.SR},
       adsurl = {https://ui.adsabs.harvard.edu/abs/2019ApJ...871..235P}
}

@ARTICLE{Ranjan2023,
       author = {{Ranjan}, Sukrit and {Schwieterman}, Edward W. and {Leung}, Michaela and {Harman}, Chester E. and {Hu}, Renyu},
        title = "{The Importance of the Upper Atmosphere to CO/O$_{2}$ Runaway on Habitable Planets Orbiting Low-mass Stars}",
      journal = {\apjl},
         year = 2023,
        month = nov,
       volume = {958},
       number = {1},
          eid = {L15},
        pages = {L15},
          doi = {10.3847/2041-8213/ad037c},
archivePrefix = {arXiv},
       eprint = {2307.08752},
 primaryClass = {astro-ph.EP},
       adsurl = {https://ui.adsabs.harvard.edu/abs/2023ApJ...958L..15R}
}

@ARTICLE{Wilson2021,
       author = {{Wilson}, David J. and {Froning}, Cynthia S. and {Duvvuri}, Girish M. and {France}, Kevin and {Youngblood}, Allison and {Schneider}, P. Christian and {Berta-Thompson}, Zachory and {Brown}, Alexander and {Buccino}, Andrea P. and {Hawley}, Suzanne and {Irwin}, Jonathan and {Kaltenegger}, Lisa and {Kowalski}, Adam and {Linsky}, Jeffrey and {Loyd}, R.~O. Parke and {Miguel}, Yamila and {Pineda}, J. Sebastian and {Redfield}, Seth and {Roberge}, Aki and {Rugheimer}, Sarah and {Tian}, Feng and {Vieytes}, Mariela},
        title = "{The Mega-MUSCLES Spectral Energy Distribution of TRAPPIST-1}",
      journal = {\apj},
         year = 2021,
        month = apr,
       volume = {911},
       number = {1},
          eid = {18},
        pages = {18},
          doi = {10.3847/1538-4357/abe771},
archivePrefix = {arXiv},
       eprint = {2102.11415},
 primaryClass = {astro-ph.SR},
       adsurl = {https://ui.adsabs.harvard.edu/abs/2021ApJ...911...18W}
}

@ARTICLE{Agol2021,
       author = {{Agol}, Eric and {Dorn}, Caroline and {Grimm}, Simon L. and {Turbet}, Martin and {Ducrot}, Elsa and {Delrez}, Laetitia and {Gillon}, Micha{\"e}l and {Demory}, Brice-Olivier and {Burdanov}, Artem and {Barkaoui}, Khalid and {Benkhaldoun}, Zouhair and {Bolmont}, Emeline and {Burgasser}, Adam and {Carey}, Sean and {de Wit}, Julien and {Fabrycky}, Daniel and {Foreman-Mackey}, Daniel and {Haldemann}, Jonas and {Hernandez}, David M. and {Ingalls}, James and {Jehin}, Emmanuel and {Langford}, Zachary and {Leconte}, J{\'e}r{\'e}my and {Lederer}, Susan M. and {Luger}, Rodrigo and {Malhotra}, Renu and {Meadows}, Victoria S. and {Morris}, Brett M. and {Pozuelos}, Francisco J. and {Queloz}, Didier and {Raymond}, Sean N. and {Selsis}, Franck and {Sestovic}, Marko and {Triaud}, Amaury H.~M.~J. and {Van Grootel}, Valerie},
        title = "{Refining the Transit-timing and Photometric Analysis of TRAPPIST-1: Masses, Radii, Densities, Dynamics, and Ephemerides}",
      journal = {\psj},
         year = 2021,
        month = feb,
       volume = {2},
       number = {1},
          eid = {1},
        pages = {1},
          doi = {10.3847/PSJ/abd022},
archivePrefix = {arXiv},
       eprint = {2010.01074},
 primaryClass = {astro-ph.EP},
       adsurl = {https://ui.adsabs.harvard.edu/abs/2021PSJ.....2....1A}
}

@ARTICLE{Batalha2019,
       author = {{Batalha}, Natasha E. and {Marley}, Mark S. and {Lewis}, Nikole K. and {Fortney}, Jonathan J.},
        title = "{Exoplanet Reflected-light Spectroscopy with PICASO}",
      journal = {\apj},
         year = 2019,
        month = jun,
       volume = {878},
       number = {1},
          eid = {70},
        pages = {70},
          doi = {10.3847/1538-4357/ab1b51},
archivePrefix = {arXiv},
       eprint = {1904.09355},
 primaryClass = {astro-ph.EP},
       adsurl = {https://ui.adsabs.harvard.edu/abs/2019ApJ...878...70B}
}

@ARTICLE{Wogan2025b,
       author = {{Wogan}, Nicholas F. and {Batalha}, Natasha E. and {Zahnle}, Kevin and {Krissansen-Totton}, Joshua and {Catling}, David C. and {Wolf}, Eric T. and {Robinson}, Tyler D. and {Meadows}, Victoria and {Arney}, Giada and {Domagal-Goldman}, Shawn},
        title = "{The Open-Source Photochem Code: A General Chemical and Climate Model for Interpreting (Exo)Planet Observations}",
      journal = {arXiv e-prints},
         year = 2025,
        month = sep,
          eid = {arXiv:2509.25578},
        pages = {arXiv:2509.25578},
          doi = {10.48550/arXiv.2509.25578},
archivePrefix = {arXiv},
       eprint = {2509.25578},
 primaryClass = {astro-ph.EP},
       adsurl = {https://ui.adsabs.harvard.edu/abs/2025arXiv250925578W}
}

@misc{Opacities2025,
  author       = {Wogan, Nicholas and
                  Lopez, Jaden and
                  Batalha, Natasha and
                  Fortney, Jonathan},
  title        = {Resampled Opacity Databases for PICASO from the Photochem model},
  month        = oct,
  year         = 2025,
  publisher    = {Zenodo},
  version      = {v1.0.0},
  doi          = {10.5281/zenodo.17381172},
  url          = {https://doi.org/10.5281/zenodo.17381172},
}

@ARTICLE{Mlawer2012,
       author = {{Mlawer}, E.~J. and {Payne}, V.~H. and {Moncet}, J.-L. and {Delamere}, J.~S. and {Alvarado}, M.~J. and {Tobin}, D.~C.},
        title = "{Development and recent evaluation of the MT\_CKD model of continuum absorption}",
      journal = {Philosophical Transactions of the Royal Society of London Series A},
         year = 2012,
        month = jun,
       volume = {370},
       number = {1968},
        pages = {2520-2556},
          doi = {10.1098/rsta.2011.0295},
       adsurl = {https://ui.adsabs.harvard.edu/abs/2012RSPTA.370.2520M}
}

@ARTICLE{Kharecha2005,
       author = {{Kharecha}, P. and {Kasting}, J. and {Siefert}, J.},
        title = "{A coupled atmosphere{\textendash}ecosystem model of the early Archean Earth}",
      journal = {Geobiology},
         year = 2005,
        month = apr,
       volume = {3},
       number = {2},
        pages = {53-76},
          doi = {10.1111/j.1472-4669.2005.00049.x},
       adsurl = {https://ui.adsabs.harvard.edu/abs/2005Gbio....3...53K}
}

@ARTICLE{Batalha2017,
       author = {{Batalha}, Natasha E. and {Mandell}, Avi and {Pontoppidan}, Klaus and {Stevenson}, Kevin B. and {Lewis}, Nikole K. and {Kalirai}, Jason and {Earl}, Nick and {Greene}, Thomas and {Albert}, Lo{\"\i}c and {Nielsen}, Louise D.},
        title = "{PandExo: A Community Tool for Transiting Exoplanet Science with JWST \& HST}",
      journal = {\pasp},
         year = 2017,
        month = jun,
       volume = {129},
       number = {976},
        pages = {064501},
          doi = {10.1088/1538-3873/aa65b0},
archivePrefix = {arXiv},
       eprint = {1702.01820},
 primaryClass = {astro-ph.IM},
       adsurl = {https://ui.adsabs.harvard.edu/abs/2017PASP..129f4501B}
}

@ARTICLE{Rathcke2025,
       author = {{Rathcke}, Alexander D. and {Buchhave}, Lars A. and {de Wit}, Julien and {Rackham}, Benjamin V. and {August}, Prune C. and {Diamond-Lowe}, Hannah and {Mendon{\c{C}}a}, Jo{\~a}o M. and {Bello-Arufe}, Aaron and {L{\'o}pez-Morales}, Mercedes and {Kitzmann}, Daniel and {Heng}, Kevin},
        title = "{Stellar Contamination Correction Using Back-to-back Transits of TRAPPIST-1 b and c}",
      journal = {\apjl},
         year = 2025,
        month = jan,
       volume = {979},
       number = {1},
          eid = {L19},
        pages = {L19},
          doi = {10.3847/2041-8213/ada5c7},
archivePrefix = {arXiv},
       eprint = {2412.16541},
 primaryClass = {astro-ph.EP},
       adsurl = {https://ui.adsabs.harvard.edu/abs/2025ApJ...979L..19R}
}

@ARTICLE{Scarsdale2024,
       author = {{Scarsdale}, Nicholas and {Wogan}, Nicholas and {Wakeford}, Hannah R. and {Wallack}, Nicole L. and {Batalha}, Natasha E. and {Alderson}, Lili and {Aguichine}, Artyom and {Wolfgang}, Angie and {Teske}, Johanna and {Moran}, Sarah E. and {L{\'o}pez-Morales}, Mercedes and {Kirk}, James and {Gordon}, Tyler and {Gao}, Peter and {Batalha}, Natalie M. and {Alam}, Munazza K. and {Adams Redai}, Jea},
        title = "{JWST COMPASS: The 3{\textendash}5 {\ensuremath{\mu}}m Transmission Spectrum of the Super-Earth L 98-59 c}",
      journal = {\aj},
         year = 2024,
        month = dec,
       volume = {168},
       number = {6},
          eid = {276},
        pages = {276},
          doi = {10.3847/1538-3881/ad73cf},
archivePrefix = {arXiv},
       eprint = {2409.07552},
 primaryClass = {astro-ph.EP},
       adsurl = {https://ui.adsabs.harvard.edu/abs/2024AJ....168..276S}
}

@ARTICLE{Espinoza2025,
       author = {{Espinoza}, N{\'e}stor and {Allen}, Natalie H. and {Glidden}, Ana and {Lewis}, Nikole K. and {Seager}, Sara and {Ca{\~n}as}, Caleb I. and {Grant}, David and {Gressier}, Am{\'e}lie and {Courreges}, Shelby and {Stevenson}, Kevin B. and {Ranjan}, Sukrit and {Col{\'o}n}, Knicole and {Morris}, Brett M. and {MacDonald}, Ryan J. and {Long}, Douglas and {Wakeford}, Hannah R. and {Valenti}, Jeff A. and {Alderson}, Lili and {Batalha}, Natasha E. and {Challener}, Ryan C. and {Huang}, Jingcheng and {Lin}, Zifan and {Louie}, Dana R. and {Mullens}, Elijah and {Valentine}, Daniel and {Mountain}, C. Matt and {Pueyo}, Laurent and {Perrin}, Marshall D. and {Bellini}, Andrea and {Kammerer}, Jens and {Libralato}, Mattia and {Rebollido}, Isabel and {Rickman}, Emily and {Sohn}, Sangmo Tony and {van der Marel}, Roeland P.},
        title = "{JWST-TST DREAMS: NIRSpec/PRISM Transmission Spectroscopy of the Habitable Zone Planet TRAPPIST-1 e}",
      journal = {\apjl},
         year = 2025,
        month = sep,
       volume = {990},
       number = {2},
          eid = {L52},
        pages = {L52},
          doi = {10.3847/2041-8213/adf42e},
archivePrefix = {arXiv},
       eprint = {2509.05414},
 primaryClass = {astro-ph.EP},
       adsurl = {https://ui.adsabs.harvard.edu/abs/2025ApJ...990L..52E}
}

@ARTICLE{Buchner2014,
       author = {{Buchner}, J. and {Georgakakis}, A. and {Nandra}, K. and {Hsu}, L. and {Rangel}, C. and {Brightman}, M. and {Merloni}, A. and {Salvato}, M. and {Donley}, J. and {Kocevski}, D.},
        title = "{X-ray spectral modelling of the AGN obscuring region in the CDFS: Bayesian model selection and catalogue}",
      journal = {\aap},
         year = 2014,
        month = apr,
       volume = {564},
          eid = {A125},
        pages = {A125},
          doi = {10.1051/0004-6361/201322971},
archivePrefix = {arXiv},
       eprint = {1402.0004},
 primaryClass = {astro-ph.HE},
       adsurl = {https://ui.adsabs.harvard.edu/abs/2014A&A...564A.125B}
}

@ARTICLE{MikalEvans2022,
       author = {{Mikal-Evans}, Thomas},
        title = "{Detecting the proposed CH$_{4}$-CO$_{2}$ biosignature pair with the James Webb Space Telescope: TRAPPIST-1e and the effect of cloud/haze}",
      journal = {\mnras},
         year = 2022,
        month = feb,
       volume = {510},
       number = {1},
        pages = {980-991},
          doi = {10.1093/mnras/stab3383},
archivePrefix = {arXiv},
       eprint = {2111.09685},
 primaryClass = {astro-ph.EP},
       adsurl = {https://ui.adsabs.harvard.edu/abs/2022MNRAS.510..980M}
}

@ARTICLE{LustigYaeger2019,
       author = {{Lustig-Yaeger}, Jacob and {Meadows}, Victoria S. and {Lincowski}, Andrew P.},
        title = "{The Detectability and Characterization of the TRAPPIST-1 Exoplanet Atmospheres with JWST}",
      journal = {\aj},
         year = 2019,
        month = jul,
       volume = {158},
       number = {1},
          eid = {27},
        pages = {27},
          doi = {10.3847/1538-3881/ab21e0},
archivePrefix = {arXiv},
       eprint = {1905.07070},
 primaryClass = {astro-ph.EP},
       adsurl = {https://ui.adsabs.harvard.edu/abs/2019AJ....158...27L}
}

@ARTICLE{Wong2022,
       author = {{Wong}, Michael L. and {Bartlett}, Stuart and {Chen}, Sihe and {Tierney}, Louisa},
        title = "{Searching for Life, Mindful of Lyfe's Possibilities}",
      journal = {Life},
         year = 2022,
        month = may,
       volume = {12},
       number = {6},
          eid = {783},
        pages = {783},
          doi = {10.3390/life12060783},
       adsurl = {https://ui.adsabs.harvard.edu/abs/2022Life...12..783W}
}

@ARTICLE{Ranjan2025,
       author = {{Ranjan}, Sukrit and {Wogan}, Nicholas F. and {Glidden}, Ana and {Wang}, Jingyu and {Stevenson}, Kevin B. and {Lewis}, Nikole and {Koskinen}, Tommi and {Seager}, Sara and {Wakeford}, Hannah R. and {van der Marel}, Roeland P.},
        title = "{The Photochemical Plausibility of Warm Exo-Titans Orbiting M Dwarf Stars}",
      journal = {\apjl},
         year = 2025,
        month = nov,
       volume = {993},
       number = {2},
          eid = {L39},
        pages = {L39},
          doi = {10.3847/2041-8213/ae1026},
archivePrefix = {arXiv},
       eprint = {2509.10611},
 primaryClass = {astro-ph.EP},
       adsurl = {https://ui.adsabs.harvard.edu/abs/2025ApJ...993L..39R}
}

@ARTICLE{Arney2026,
       author = {{Arney}, Giada and {Parenteau}, Niki and {Hinkel}, Natalie and {Mamajek}, Eric and {Krissansen-Totton}, Joshua and {Olson}, Stephanie and {Schwieterman}, Edward and {Walker}, Sara and {Fogarty}, Kevin and {Kopparapu}, Ravi and {Lustig-Yaeger}, Jacob and {Moussa}, Mark and {Ranjan}, Sukrit and {Singh}, Garima and {Sousa-Silva}, Clara and {Belikov}, Ruslan and {Frissell}, Maxwell and {Gilbert-Janziek}, Samantha and {Kofman}, Vincent and {Latouf}, Natasha and {Limbach}, Mary Anne and {Morgan}, Rhonda and {Stark}, Christopher and {Tokadjian}, Armen and {Ulses}, Anna Grace and {Wogan}, Nicholas and {Wong}, Mike and {Young}, Amber},
        title = "{Habitable Worlds Observatory (HWO): Living Worlds Community Working Group: The Search for Life on Potentially Habitable Exoplanets}",
      journal = {arXiv e-prints},
         year = 2026,
        month = jan,
          eid = {arXiv:2601.09766},
        pages = {arXiv:2601.09766},
          doi = {10.48550/arXiv.2601.09766},
archivePrefix = {arXiv},
       eprint = {2601.09766},
 primaryClass = {astro-ph.IM},
       adsurl = {https://ui.adsabs.harvard.edu/abs/2026arXiv260109766A}
}

@ARTICLE{Parenteau2026,
       author = {{Parenteau}, Niki and {Ulses}, Anna Grace and {Metz}, Connor and {Kiang}, Nancy Y. and {Coelho}, Ligia F. and {Schwieterman}, Edward and {Grone}, Jonathan and {Roccetti}, Giulia and {Berdyugina}, Svetlana and {Alei}, Eleonora and {Patty}, Lucas and {Lafleche}, Emilie and {Matsuo}, Taro and {Cardace}, Dawn and {Borges}, Schuyler and {Mandel}, Avi and {Gordon}, Kenneth and {Krissansen-Totton}, Joshua and {Arney}, Giada},
        title = "{Habitable Worlds Observatory Living Worlds Working Group: Surface Biosignatures on Potentially Habitable Exoplanets}",
      journal = {arXiv e-prints},
         year = 2026,
        month = jan,
          eid = {arXiv:2601.08883},
        pages = {arXiv:2601.08883},
          doi = {10.48550/arXiv.2601.08883},
archivePrefix = {arXiv},
       eprint = {2601.08883},
 primaryClass = {astro-ph.IM},
       adsurl = {https://ui.adsabs.harvard.edu/abs/2026arXiv260108883P}
}

@ARTICLE{Guzman2013,
       author = {{Guzm{\'a}n-Marmolejo}, Andr{\'e}s and {Segura}, Ant{\'\i}gona and {Escobar-Briones}, Elva},
        title = "{Abiotic Production of Methane in Terrestrial Planets full access}",
      journal = {Astrobiology},
         year = 2013,
        month = jun,
       volume = {13},
       number = {6},
        pages = {550-559},
          doi = {10.1089/ast.2012.0817},
       adsurl = {https://ui.adsabs.harvard.edu/abs/2013AsBio..13..550G}
}

@ARTICLE{Schindler2000,
       author = {{Schindler}, Trent L. and {Kasting}, James F.},
        title = "{Synthetic Spectra of Simulated Terrestrial Atmospheres Containing Possible Biomarker Gases}",
      journal = {\icarus},
         year = 2000,
        month = may,
       volume = {145},
       number = {1},
        pages = {262-271},
          doi = {10.1006/icar.2000.6340},
       adsurl = {https://ui.adsabs.harvard.edu/abs/2000Icar..145..262S}
}

@ARTICLE{Allen2026,
       author = {{Allen}, Natalie H. and {Espinoza}, N{\'e}stor and {Boehm}, V.~A. and {Ca{\~n}as}, Caleb I. and {Stevenson}, Kevin B. and {Lewis}, Nikole K. and {MacDonald}, Ryan J. and {Morris}, Brett M. and {Agol}, Eric and {Col{\'o}n}, Knicole and {Diamond-Lowe}, Hannah and {Glidden}, Ana and {Gressier}, Am{\'e}lie and {Huang}, Jingcheng and {Lin}, Zifan and {Long}, Douglas and {Louie}, Dana R. and {MacGregor}, Meredith A. and {Pueyo}, Laurent and {Rackham}, Benjamin V. and {Ranjan}, Sukrit and {Seager}, Sara and {Tovar Mendoza}, Guadalupe and {Valenti}, Jeff A. and {Valentine}, Daniel and {van der Marel}, Roeland P. and {Wakeford}, Hannah R.},
        title = "{JWST TRAPPIST-1 e/b Program: Motivation and First Observations}",
      journal = {\aj},
         year = 2026,
        month = feb,
       volume = {171},
       number = {2},
          eid = {105},
        pages = {105},
          doi = {10.3847/1538-3881/ae28cb},
archivePrefix = {arXiv},
       eprint = {2512.07695},
 primaryClass = {astro-ph.EP},
       adsurl = {https://ui.adsabs.harvard.edu/abs/2026AJ....171..105A}
}

@ARTICLE{Gordon2017,
       author = {{Gordon}, I.~E. and {Rothman}, L.~S. and {Hill}, C. and {Kochanov}, R.~V. and {Tan}, Y. and {Bernath}, P.~F. and {Birk}, M. and {Boudon}, V. and {Campargue}, A. and {Chance}, K.~V. and {Drouin}, B.~J. and {Flaud}, J. -M. and {Gamache}, R.~R. and {Hodges}, J.~T. and {Jacquemart}, D. and {Perevalov}, V.~I. and {Perrin}, A. and {Shine}, K.~P. and {Smith}, M. -A.~H. and {Tennyson}, J. and {Toon}, G.~C. and {Tran}, H. and {Tyuterev}, V.~G. and {Barbe}, A. and {Cs{\'a}sz{\'a}r}, A.~G. and {Devi}, V.~M. and {Furtenbacher}, T. and {Harrison}, J.~J. and {Hartmann}, J. -M. and {Jolly}, A. and {Johnson}, T.~J. and {Karman}, T. and {Kleiner}, I. and {Kyuberis}, A.~A. and {Loos}, J. and {Lyulin}, O.~M. and {Massie}, S.~T. and {Mikhailenko}, S.~N. and {Moazzen-Ahmadi}, N. and {M{\"u}ller}, H.~S.~P. and {Naumenko}, O.~V. and {Nikitin}, A.~V. and {Polyansky}, O.~L. and {Rey}, M. and {Rotger}, M. and {Sharpe}, S.~W. and {Sung}, K. and {Starikova}, E. and {Tashkun}, S.~A. and {Auwera}, J. Vander and {Wagner}, G. and {Wilzewski}, J. and {Wcis{\l}o}, P. and {Yu}, S. and {Zak}, E.~J.},
        title = "{The HITRAN2016 molecular spectroscopic database}",
      journal = {\jqsrt},
         year = 2017,
        month = dec,
       volume = {203},
        pages = {3-69},
          doi = {10.1016/j.jqsrt.2017.06.038},
       adsurl = {https://ui.adsabs.harvard.edu/abs/2017JQSRT.203....3G}
}

@ARTICLE{Hargreaves2020,
       author = {{Hargreaves}, Robert J. and {Gordon}, Iouli E. and {Rey}, Michael and {Nikitin}, Andrei V. and {Tyuterev}, Vladimir G. and {Kochanov}, Roman V. and {Rothman}, Laurence S.},
        title = "{An Accurate, Extensive, and Practical Line List of Methane for the HITEMP Database}",
      journal = {\apjs},
         year = 2020,
        month = apr,
       volume = {247},
       number = {2},
          eid = {55},
        pages = {55},
          doi = {10.3847/1538-4365/ab7a1a},
archivePrefix = {arXiv},
       eprint = {2001.05037},
 primaryClass = {astro-ph.EP},
       adsurl = {https://ui.adsabs.harvard.edu/abs/2020ApJS..247...55H}
}

@ARTICLE{Currie2025,
       author = {{Currie}, Miles H. and {Meadows}, Victoria S.},
        title = "{There's More to Life in Reflected Light: Simulating the Detectability of a Range of Molecules for High-contrast, High-resolution Observations of Nontransiting Terrestrial Exoplanets}",
      journal = {\psj},
         year = 2025,
        month = apr,
       volume = {6},
       number = {4},
          eid = {96},
        pages = {96},
          doi = {10.3847/PSJ/adc004},
archivePrefix = {arXiv},
       eprint = {2503.08592},
 primaryClass = {astro-ph.EP},
       adsurl = {https://ui.adsabs.harvard.edu/abs/2025PSJ.....6...96C}
}

@unpublished{sneedprep,
    author = {{Sneed}, Evan L. and {Schwieterman}, Edward W. and {Peacock}, Sarah R. and {Leung}, Michaela and {Wogan}, Nicholas F. and {Lyons}, Timothy W.},
    title = "{Ultraviolet-Driven Atmospheric Degeneracies Could Challenge Conventional Biosignature Frameworks for an Archean-Analog TRAPPIST-1}",
    pubstate = {inpreparation},
    year=2026,
    note={submitting soon!}
}

@ARTICLE{Quanz2022,
       author = {{Quanz}, S.~P. and {Ottiger}, M. and {Fontanet}, E. and {Kammerer}, J. and {Menti}, F. and {Dannert}, F. and {Gheorghe}, A. and {Absil}, O. and {Airapetian}, V.~S. and {Alei}, E. and {Allart}, R. and {Angerhausen}, D. and {Blumenthal}, S. and {Buchhave}, L.~A. and {Cabrera}, J. and {Carri{\'o}n-Gonz{\'a}lez}, {\'O}. and {Chauvin}, G. and {Danchi}, W.~C. and {Dandumont}, C. and {Defr{\'e}re}, D. and {Dorn}, C. and {Ehrenreich}, D. and {Ertel}, S. and {Fridlund}, M. and {Garc{\'\i}a Mu{\~n}oz}, A. and {Gasc{\'o}n}, C. and {Girard}, J.~H. and {Glauser}, A. and {Grenfell}, J.~L. and {Guidi}, G. and {Hagelberg}, J. and {Helled}, R. and {Ireland}, M.~J. and {Janson}, M. and {Kopparapu}, R.~K. and {Korth}, J. and {Kozakis}, T. and {Kraus}, S. and {L{\'e}ger}, A. and {Leedj{\"a}rv}, L. and {Lichtenberg}, T. and {Lillo-Box}, J. and {Linz}, H. and {Liseau}, R. and {Loicq}, J. and {Mahendra}, V. and {Malbet}, F. and {Mathew}, J. and {Mennesson}, B. and {Meyer}, M.~R. and {Mishra}, L. and {Molaverdikhani}, K. and {Noack}, L. and {Oza}, A.~V. and {Pall{\'e}}, E. and {Parviainen}, H. and {Quirrenbach}, A. and {Rauer}, H. and {Ribas}, I. and {Rice}, M. and {Romagnolo}, A. and {Rugheimer}, S. and {Schwieterman}, E.~W. and {Serabyn}, E. and {Sharma}, S. and {Stassun}, K.~G. and {Szul{\'a}gyi}, J. and {Wang}, H.~S. and {Wunderlich}, F. and {Wyatt}, M.~C. and {LIFE Collaboration}},
        title = "{Large Interferometer For Exoplanets (LIFE). I. Improved exoplanet detection yield estimates for a large mid-infrared space-interferometer mission}",
      journal = {\aap},
         year = 2022,
        month = aug,
       volume = {664},
          eid = {A21},
        pages = {A21},
          doi = {10.1051/0004-6361/202140366},
archivePrefix = {arXiv},
       eprint = {2101.07500},
 primaryClass = {astro-ph.EP},
       adsurl = {https://ui.adsabs.harvard.edu/abs/2022A&A...664A..21Q}
}

@ARTICLE{Wogan2024,
       author = {{Wogan}, Nicholas F. and {Batalha}, Natasha E. and {Zahnle}, Kevin J. and {Krissansen-Totton}, Joshua and {Tsai}, Shang-Min and {Hu}, Renyu},
        title = "{JWST Observations of K2-18b Can Be Explained by a Gas-rich Mini-Neptune with No Habitable Surface}",
      journal = {\apjl},
         year = 2024,
        month = mar,
       volume = {963},
       number = {1},
          eid = {L7},
        pages = {L7},
          doi = {10.3847/2041-8213/ad2616},
archivePrefix = {arXiv},
       eprint = {2401.11082},
 primaryClass = {astro-ph.EP},
       adsurl = {https://ui.adsabs.harvard.edu/abs/2024ApJ...963L...7W}
}
\bibliographystyle{aasjournalv7}

\end{document}